\documentclass[12pt,reqno,a4paper]{article}
\usepackage{authblk}
\usepackage[margin=1in]{geometry}
\RequirePackage{amsthm,amsmath,amsfonts,amssymb}
\RequirePackage{natbib}
\RequirePackage[colorlinks,citecolor=blue,urlcolor=blue]{hyperref}
\usepackage{booktabs,longtable}
\usepackage{graphicx}
\usepackage{xcolor}
\usepackage{epstopdf}
\usepackage{bm}
\usepackage[linesnumbered,ruled,vlined]{algorithm2e}
\usepackage{multirow}
\usepackage{setspace}
\usepackage{tikz}

\newtheorem{proposition}{Proposition}
\newcommand{\bmX}{\bm{X}}
\newcommand{\bmpi}{\bm{\pi}}
\newcommand{\given}{\, \vert \,}
\newcommand{\bmB}{\bm{B}}
\newcommand{\bmZ}{\bm{Z}}
\newcommand{\bmalpha}{\bm{\alpha}}
\newcommand{\bmbeta}{\bm{\beta}}
\newcommand{\bmR}{\bm{R}}
\newcommand{\itoj}{i \rightarrow j}
\newcommand{\ifromj}{i \leftarrow j}
\newcommand{\bmc}{\bm{c}}
\newcommand{\bmmu}{\bm{\mu}}
\newcommand{\bmTheta}{\bm{\Theta}}
\newcommand{\bmgamma}{\bm{\gamma}}
\newcommand{\bmphi}{\bm{\phi}}
\newcommand{\bmx}{\bm{x}}
\DeclareMathOperator*{\argmax}{arg\,max}

\graphicspath{{./}{../figures/}}

\newcommand{\blue}[1]{\textcolor{blue}{#1}}

\title{Variational Bayesian Inference for Bipartite Mixed-membership 
Stochastic Block Model with Applications to Collaborative Filtering}

\author[1]{Jie Liu}

\author[1]{Zifeng Ye}

\affil[1]{Department of Statistics and Finance, University of 
Science and Technology of China, Hefei, Anhui 230026, China}

\author[2]{Kun Chen}

\affil[2]{Department of Statistics, University of Connecticut, 
	Storrs, CT 06269, USA}

\author[3,4,*]{Panpan Zhang}

\affil[3]{Department of Biostatistics, Vanderbilt University Medical 
	Center, Nashville, TN 37203, USA}

\affil[4]{Vanderbilt Memory \& Alzheimer's Center, Nashville, TN 
	37212, USA}

\affil[*]{Correspondence: 
\href{mailto:panpan.zhang@vumc.org}{panpan.zhang@vumc.org}}

\begin{document}

\maketitle

\begin{abstract}
	Motivated by the connections between collaborative filtering and 
	network
	clustering, we consider a network-based approach to improving 
	rating prediction in recommender systems. We propose a novel 
	Bipartite Mixed-Membership Stochastic Block Model 
	($\mathrm{BM}^2$) with a conjugate prior from the exponential 
	family. We derive the analytical expression of the model and 
	introduce a variational Bayesian expectation-maximization 
	algorithm, which is computationally feasible for approximating 
	the untractable posterior distribution. We carry out extensive 
	simulations to show that $\mathrm{BM}^2$ provides more accurate 
	inference than standard SBM with the emergence of outliers. 
	Finally, we apply the proposed model to a MovieLens dataset, and 
	find that it outperforms other competing methods for 
	collaborative filtering.
\end{abstract}

	
\section{Introduction}
\label{sec:intro}

The problem of information overload primarily arises from easy
access to a huge volume of information. A {\em recommender system}
\citep[RS,][]{Aggarwal2016recommender} is a class of information
filtering system that predicts the rating of an item given by a
user. Recommender systems play a critical role in the face of
excessive information about an increasing number of users' interests 
on an ever-growing list of items, as it helps filter out irrelevant 
information accurately and efficiently.
Recommender systems have found a plethora of applications, such as 
movies~\citep{Bell2007scalable, Harper2015movielens},
books~\citep{Linden2003amazon}, and Jester 
jokes~\citep{Goldberg2001eigentaste}. In general, there are three 
classes of strategies of RS, namely, content-based methods, {\em 
collaborative filtering}, and hybrid methods. In the present study, 
we focus on collaborative filtering, which refers to the method of 
multiple users sharing the recommendations in the form of ratings 
under the assumption that users from the same group give similar 
ratings to similar items. Collaborative filtering effectively 
bridges the
user space and the item space, but does not require the content
analysis of the item, rendering it a versatile and widely used
algorithm for recommender system studies. For example, 
\citet{Jamali2009using} developed an approach to incorporating a 
social network into a top-$N$ recommender system based on nearest 
neighbors. \citet{Liu2013bayesian} proposed a Bayesian probabilistic 
matrix factorization algorithm that was applied to trust-aware 
recommender systems for large datasets. We refer the interested 
readers to \citet{Yang2014survey} for a succinct review of 
collaborative filtering.

Stochastic Block Models \citep[SBMs,][]{Snijders1997estimation} 
are a class of widely used models for detecting latent block 
structures (i.e., communities) in networks. The underlying principle 
of SBMs is to group/cluster the nodes presenting similar 
characteristics or features together so as to recover the community 
structure of the network. With the increase of complexity of 
network-based systems, a number of SBM variants have been developed 
recently. \citet{Airoldi2008mixed} proposed a mixed-membership SBM 
allowing for fuzzy clustering, i.e., each node may belong to 
multiple communities. 
\citet{Karrer2011stochastic} introduced a class of degree-corrected 
SBMs by accounting for node-specific heterogeneity. More recently, 
\citet{Bouveyron2018stochastic}
integrated SBM and Latent Dirichlet 
Allocation~\citep{Blei2003latent}, a generative probabilistic 
model, for community detection in networks with textual
edges (e.g., social media data). See \citet{Lee2019review} for a 
comprehensive review of SBM and its variants with applications to 
network clustering.

Realizing the connections and similarities between the setups of 
collaborative filtering and network clustering, we aim to 
utilize SBM to improve the accuracy of rating predictions for 
collaborative filtering. Different from classical collaborative 
filtering methods like user-based or item-based methods undergoing 
the limitations such as incapability of handling sparse data, lack 
of scalability, and requiring large storage space, we consider a 
model-based approach utilizing state-to-art network analysis tools. 
More precisely, we model a recommender system of user and item 
spaces as a {\em bipartite network} consisting of user and item 
nodes. The {\em adjacency matrix} of the network is represented by 
the rating matrix, where the $(i, j)$-th entry records the rating of 
item $j$ by user $i$. Since SBM assumes that the probability 
distribution of edges in a network is governed by the communities to 
which the nodes belong, community detection algorithms for SBMs can 
be utilized to search for the groups of users who tend to give 
similar ratings to the items from the same group. Consequently,
predicting the unobserved rates in the rating matrix is analogous to 
link prediction in network analysis.

To the best of our knowledge, there is limited work on rating 
prediction via network modeling. \citet{Guimera2012predicting} 
developed an ensemble of SBM to predict the individuals' preferences 
in a Bayesian framework. However, there is a lack of explicit 
specification of the data generating process, making their model 
difficult to interpret. \citet{Godoy2016accurate} adopted a 
mixed-membership stochastic block model (MMSBM) to predict the 
unobserved rates by the active users from a large-scale dataset, and 
proposed an {\em expectation-maximization} (EM) algorithm for 
inference. Later on, \citet{Godoy2019network} applied the model to 
filtering and provided some personalized advice to the users. 
However, the (joint) user-item clusters that they reported fail to 
uncover the respective membership information for users and items, 
and their inference was lacking clear statistical interpretation. 

Motivated by the limitations of existing research, we propose a 
novel Bipartite Mixed-Membership Stochastic Block Model 
($\mathrm{BM}^2$) endowed with explicit generative processes for 
bipartite networks. One of the main improvements of the proposed 
model (compared to the MMSBM by~\citet{Godoy2016accurate}) is that 
it is capable of capturing directed edges emanating out of both node 
sets, making it applicable to not only the motivating (undirected) 
network data in the present study but also more general directed 
bipartite networks. The inference of the proposed model is primarily 
made in a Bayesian framework, which has become popular for network 
model analysis due to its statistical rigor and overall good 
performance~\citep{Guimera2012predicting, Jorgensen2016bayesian, 
	Peng2016bayesian}. $\mathrm{BM}^2$ inherits the feature of SBM 
	that 
users and items may belong to multiple clusters simultaneously, and 
is able to clearly specify the mixed membership information (in 
terms of a vector) for each user and item. Besides, its underlying 
structure coincides with the collaborative filtering assumption that 
users from the same group rate similar items similarly. By selecting 
appropriate conjugate prior distributions, we are able to derive the 
probability distribution of each parameter explicitly, rendering 
more practical interpretations. To circumvent the challenge of 
computational intractability for maximizing the posterior 
distributions, we develop a variational EM algorithm that 
efficiently approximates the posterior distributions and provide 
accurate inference~\citep{Airoldi2008mixed, Daudin2008mixture, 
	Blei2017variational}. Through extensive simulations, we show 
	that 
$\mathrm{BM}^2$ is robust in the presence of outliers 
and the variational EM algorithm is scalable. Additionally, we 
demonstrate that the proposed model outperforms non-Bayesian 
SBM~\citep{Godoy2016accurate} and several other competing methods in 
an application to a MovieLens dataset.

The rest of the paper is organized as follows. We review some
fundamental properties of SBM in Section~\ref{sec:pre}, and 
propose the Bipartite Mixed-membership Stochastic Block Model 
($\mathrm{BM}^2$) in Section~\ref{sec:BM2}, where the explicit 
derivations of the likelihood are provided. In Section~\ref{sec:EM}, 
we investigate the variational inference of the proposed model and 
introduce a 
variational EM algorithm. We carry out an extensive simulation study 
in Section~\ref{sec:sim}, followed by an application to the 
MovieLens dataset in 
Section~\ref{sec:movelens}. Some concluding remarks and future 
research directions are provided in Section~\ref{sec:dis}.

\section{Preliminaries}
\label{sec:pre}

Standard stochastic block model
\citep[SBM,][]{Snijders1997estimation} considers an
unweighted, undirected network $G(V, E)$ consisting of $|V| = N$
nodes and $|E|$ edges, where $V$ and $E$ respectively denote the 
node and edge sets, and $|V|$ represents the cardinality of set $V$.
The structure of $G(V, E)$ is represented by a binary, symmetric
adjacency matrix $\bmX := (X_{i,j})_{N \times N}$, where $X_{ij} 
=
1$ if the node labeled with $i$ is connected with the node 
labeled
with $j$; $X_{ij} = 0$, otherwise. Assume that the $N$ nodes from
$G(V, E)$ are partitioned into $K \in \mathbb{N}$ clusters. Let
$\bmpi_i := (\pi_{i1}, \pi_{i2}, \ldots, \pi_{iK})$ subject to
$\sum_{k = 1}^{K} \pi_{ik} = 1$ be a latent variable for each $i 
\in
V$, where $\pi_{ik}$, $k = 1, 2, \ldots, K$, denotes the
probability that node $i$ belongs to cluster $k$. Moreover, for 
each
$i \in V$, there exists a variable $\bmZ_i$ indicating the 
cluster
of $i$. It is conventional to assume
\[\bmZ_i \sim {\rm Multinomial} (1 ; \bmpi_i).\]

Under the assumption that the connectivity between any pair of 
nodes
is completely governed by the clusters that they are respectively
assigned to. Let $\bmB := (B_{kl})_{K \times K}$ be a block-wise
matrix, where $B_{kl}$, $k, l = 1, 2, \ldots, K$, represents the
link probability between cluster $k$ and cluster $l$. In 
addition,
let $\bmZ = (\bmZ_1, \bmZ_2, \ldots, \bmZ_n)$ be the collection
of all $\bmZ_i$'s. SBM assumes an independent Bernoulli model
conditional on $\bmZ$; that is
\begin{align*}
	p(\bmX \given \bmZ, \bmB, \bmpi) &= \prod_{i < j} p(X_{ij}
	\given \bmZ_i, \bmZ_j, \bmB) \prod_{i = 1}^{N} p(\bmZ_i 
	\given
	\bmpi_i)
	\\ &= \prod_{i < j} {\rm Bernoulli}(\bmZ_i^{\top} \bmB 
	\bmZ_j)
	\prod_{i = 1}^{N} p(\bmZ_i \given \bmpi_i).
\end{align*}
The
inference of SBM is usually done in a Bayesian framework. For
instance, \citet{Nowicki2001estimation} derived the posterior
estimates of SBM through an algorithm based on Gibbs sampling;
\citet{Airoldi2008mixed} investigated the posterior inference of
MMSBM via a variational Bayesian EM algorithm;
\citet{Ouyang2018model} developed a Markov Chain Monte Carlo 
(MCMC)
algorithm integrating the Metropolis-Hastings algorithm and a 
Gibbs
sampler to explore an extended SBM.

\section{Bipartite Mixed-membership Stochastic Block Model}
\label{sec:BM2}

We propose a novel Bipartite Mixed-Membership Stochastic
Block Model ($\mathrm{BM}^2$) for recommender system in this 
section. 
Section~\ref{sec:model} provides the
detailed descriptions of the proposed model, and
Section~\ref{sec:likelihood} derives the likelihood of the model
explicitly.

\subsection{Model Description}
\label{sec:model}

Let us consider a recommender system comprised of $N$ users and $M$ 
items, which 
are
assumed to be partitioned into $K$ and $L$ clusters, 
respectively.
Each user $i$ is associated with a latent probability variable
$\bmpi_{i}^{U}$ of length $K$, where the $k$-th component
$\pi_{ik}^U$ is the probability that user $i$ is categorized into
cluster $k$, for $k = 1, 2, \ldots K$. By convention, we have
$\pi_{ik}^{U} \ge 0$ for all $k$ and $\sum_{k = 1}^{K} 
\pi_{ik}^{U}
= 1$. Analogously, for each item~$j$, there is a latent 
probability
variable $\bmpi_{j}^{I}$, which is defined in a similar
manner as $\bmpi_{i}^{U}$. We use superscripts $U$ and $I$ to
distinguish the probability vectors for users and items. 
According
to the natural characteristics of $\bmpi_i^{U}$ and 
$\bmpi_j^{I}$,
we assume that $\bmpi_i^{U}$'s and $\bmpi_i^{I}$'s are
independently following the Dirichlet distributions with
hyperparameters $\bmalpha$ and $\bmbeta$, respectively, i.e.,
\begin{align*}
	\bmpi_{i}^{U} &\sim {\rm Dirichlet}(\bmalpha), i = 1, 2, 
	\ldots,
	N;
	\\ \bmpi_{j}^{I} &\sim {\rm Dirichlet}(\bmbeta), j = 1, 2,
	\ldots, M.
\end{align*}

The network associated with the recommender system is bipartite, 
since there is 
no
interactions among the users or items. Since we have two node
spaces, we introduce two kinds of indicator vectors respectively
defined as follows to capture the edge directions and weights in 
the
model. Specifically, let $\bmZ_{\itoj}^{U} := (Z_{\itoj, 
	k}^{U})_{k
	= 1}^{K}$ be the membership indicator for the user space, 
where
$Z_{\itoj, k}^{U} = 1$ indicates that user $i$ is from user 
cluster
$k$ when he/she rates item $j$. Vice versa, in the membership 
indicator
vector for the item space $\bmZ_{\ifromj}^{I} := (Z_{\ifromj,
	l}^{I})_{l = 1}^{L}$, where $Z_{\ifromj, l}^{I} = 1$ 
indicates that
item $j$ is from item cluster $l$ when it gets a rating from user
$i$. Similar to MMSBM, for each pair of user $i$ and item $j$, we
assume
\begin{align*}
	\bmZ_{\itoj}^{U} &\sim {\rm Multinomial}(\pi_i^{U};1),
	\\ \bmZ_{\ifromj}^{I} &\sim {\rm Multinomial}(\pi_j^{I};1).
\end{align*}

We use $\bmR := (R_{ij})_{N \times M}$ to record the item ratings
given by the users. For the sake of practicality, we assume that 
the
ratings only take values from a finite discrete set, say
$\mathcal{C}$. Let $S = |\mathcal{C}|$ be the number of elements 
in
set $\mathcal{C}$. Without loss of generality, let the distinct
values from $\mathcal{C}$ be ordered from the smallest to the
largest, i.e., $C_1 < C_2 < \cdots < C_S$. For each $R_{ij}$, the
event $\{R_{ij} = C_s\}$ for $s = 1, 2, \ldots, S$, is 
represented
by a user-item-based indicator vector $\bmc_{ij} := (c_{ij, 1},
c_{ij, 2}, \ldots, c_{ij, S})$, in which $c_{ij, s} = 1$, whereas
the rest are equal to $0$. Meanwhile, for each rating $C_s$, we
define a $K \times L$ matrix $\bmmu_s := (\mu_{kl, s})$ 
(analogous
to $\bmB$ for standard SBM) to depict the block level linking
probability distribution. For convenience, let $\bmmu$ be the 
array
collecting $\bmmu_1, \bmmu_2, \ldots, \bmmu_S$. Moreover, let
$\bmZ_{\rightarrow}^{U}$ and $\bmZ_{\leftarrow}^{I}$ denote the
collections of all $\bmZ_{\itoj}^{U}$'s and 
$\bmZ_{\ifromj}^{I}$'s,
respectively. Conditional on $\bmmu$, $\bmZ_{\rightarrow}^{U}$ 
and
$\bmZ_{\leftarrow}^{I}$, the $\mathrm{BM}^2$ is given by
\begin{align*}
	p(\bmR \given \bmZ_{\rightarrow}^{U}, \bmZ_{\leftarrow}^{I},
	\bmmu) &= \prod_{i, j} \prod_{s = 1}^{S} p(R_{ij} = C_s 
	\given
	\bmZ_{\itoj}^{U}, \bmZ_{\ifromj}^{I}, \bmmu_s)
	\\ &= \prod_{i, j} \prod_{k = 1}^{K} \prod_{l = 1}^{L}
	\left(\prod_{s = 1}^{S} \mu_{kl, s}^{c_{ij,
			s}}\right)^{Z_{\itoj, k}^{U} Z_{\ifromj, l}^{I}}
\end{align*}

\subsection{Likelihood Derivation}
\label{sec:likelihood}

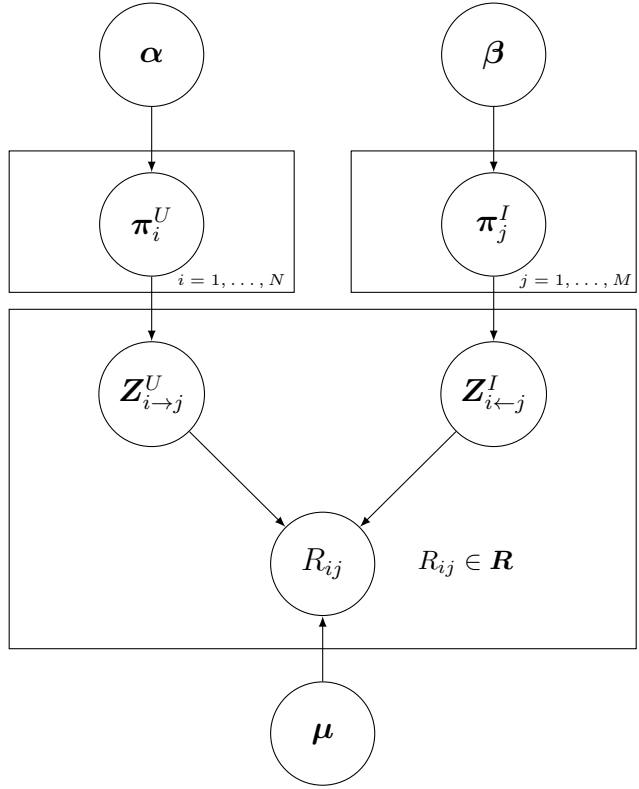
\begin{figure}[tbp]
	\begin{center}
		\begin{tikzpicture}[scale = 0.75]
			\node[draw, circle, minimum size = 1.37cm] at (0, 0)
			(alpha) {$\bmalpha$};
			\node[draw, circle, minimum size = 1.37cm] at (6, 0)
			(beta) {$\bmbeta$};
			\node[draw, circle, minimum size = 1.37cm] at (0, -3)
			(pii) {$\bmpi_i^{U}$};
			\node[draw, circle, minimum size = 1.37cm] at (6, -3)
			(pij) {$\bmpi_j^{I}$};
			\draw[-latex] (alpha)--(pii);
			\draw[-latex] (beta)--(pij);
			\node[draw, circle, minimum size = 1.37cm] at (0, -6)
			(zi) {$\bmZ_{\itoj}^{U}$};
			\node[draw, circle, minimum size = 1.37cm] at (6, -6)
			(zj) {$\bmZ_{\ifromj}^{I}$};
			\draw[-latex] (pii)--(zi);
			\draw[-latex] (pij)--(zj);
			\node[draw, circle, minimum size = 1.37cm] at (3, -9)
			(Rij) {$R_{ij}$};
			\draw[-latex] (zi)--(Rij);
			\draw[-latex] (zj)--(Rij);
			\node[draw, circle, minimum size = 1.37cm] at (3, 
			-12)
			(mu) {$\bmmu$};
			\draw[-latex] (mu)--(Rij);
			\draw (-2.5, -1.7) rectangle (2.5, -4.2);
			\draw (3.5, -1.7) rectangle (8.5, -4.2);
			\draw (-2.5, -4.5) rectangle (8.5, -10.5);
			\node[text width = 2cm] at (1.6, -4)
			{{\fontsize{6}{6} $i = 1, \ldots, N$}};
			\node[text width = 2cm] at (7.6, -4)
			{{\fontsize{6}{6} $j = 1, \ldots, M$}};
			\node[text width = 2cm] at (6, -9) {\footnotesize
				$R_{ij} \in \bmR$};
		\end{tikzpicture}
	\end{center}
	\caption{Diagram for the generating process of the proposed
		model.}
	\label{fig:diagram}
\end{figure}

We illustrate the explicit generating process for 
$\mathrm{BM}^2$, and
then derive the likelihood function of the model. The generating
process of the model proposed in 
Section~\ref{sec:BM2} is
presented as a three-step procedure as follows, accompanied by a 
graphical illustration in Figure~\ref{fig:diagram}. 
\begin{enumerate}
	\item Given hyperparameters $\bmalpha$ and $\bmbeta$, for 
	each
	user $i \in \left\{1, 2, \ldots, N\right\}$ and
	item $j \in \left\{1, 2, \ldots, M\right\}$,
	\begin{enumerate}
		\item draw a $K$-dimensional latent probability variable
		$\bmpi_{i}^{U} \sim {\rm Dirichlet}(\bmalpha)$;
		\item draw an $L$-dimensional latent probability variable
		$\bmpi_{j}^{I} \sim {\rm Dirichlet}(\bmbeta)$.
	\end{enumerate}
	\item For each observed rating $R_{ij} \in \bmR$,
	\begin{enumerate}
		\item draw a $K$-dimensional membership indicator vector
		$\bmZ_{\itoj}^{U} \sim {\rm Multinomial}(\bmpi_{i}^{U};1)$;
		\item draw an $L$-dimensional membership indicator vector
		$\bmZ_{\ifromj}^{I} \sim {\rm 
			Multinomial}(\bmpi_{j}^{I};1)$.
	\end{enumerate}
	\item For each observed rating $R_{ij} \in \bmR$, given
	$Z_{\itoj, k}^{U} = 1$ and $Z_{\ifromj, l}^{I} = 1$, draw a
	rating indicator vector $\bmc_{ij} \sim {\rm
		Multinomial}(\bmmu_{kl};1)$, where $\bmmu_{kl} := 
	(\mu_{kl, s})_{s
		= 1}^{S}$, for which we have $\sum_{s = 1}^{S} \mu_{kl, 
		s} =
	1$.	
\end{enumerate}

To further
simplify the notations, let $\bmpi$ be the collection of
$\bmpi_{i}^{U}$ for all $i$ and $\bmpi_{j}^{I}$ for all $j$, and 
let
$\bmZ$ contain the membership indicators $\bmZ_{\itoj}^{U}$ and
$\bmZ_{\ifromj}^{I}$ for all $i$ and $j$. In what follows, the
$\mathrm{BM}^2$ conditional on hyperparameter set $\bmTheta = 
\{\bmalpha,
\bmbeta\}$ and the array consisting of block matrices $\bmmu$ is 
a
multilevel model given by
\begin{align} 
	p(\bmR, \bmpi, \bmZ \given \bmTheta, \bmmu) &= \prod_{i, j}
	p(\bmc_{ij} \given \bmZ_{\itoj}^{U}, \bmZ_{\ifromj}^{I},
	\bmmu)p(\bmZ_{\itoj}^{U} \given 
	\bmpi_i^{U})p(\bmZ_{\ifromj}^{I}
	\given \bmpi_j^{I}) \nonumber
	\\ \label{eq:model} &\qquad{}\times \prod_{i =
		1}^{N}p(\bmpi_i^{U} \given
	\bmalpha)\prod_{j = 1}^{M}p(\bmpi_j^{I} \given \bmbeta),
\end{align}
which is called the complete data likelihood function. Consequently,
the prediction of missing ratings in $\bmR$ becomes a
standard inference problem. We would like to point out that, 
unlike~\citet{Tan2016topic} who imposed a beta prior to the entries 
in $\bmmu$, we do not consider any prior for $\bmmu$ as we do not 
see significant improvement in prediction accuracy through 
simulations and meanwhile it potentially avoids the issue of model 
overfitting. 

For the recommender system, we have two different node types, i.e., 
users and items, and each type of nodes is assumed to have 
its
own clustering mechanism. Accordingly, we introduce $\bmpi_i^{U}$
and $\bmpi_j^{I}$ to the proposed model. Moreover, we use 
Dirichlet
distributions as priors for $\bmpi_i^{U}$ and $\bmpi_j^{I}$ such
that all possibilities are considered, even for extreme scenarios
which are penalized with low probabilities. Conditional on
membership indicators $\bmZ$ and block level structure $\bmmu$, 
the
distribution of each of the ratings in $\bmR$ follows a 
multinomial
distribution, which is suitable for recommender system.

\section{Variational EM Algorithm}
\label{sec:EM}

To predict the unobserved rates via the proposed model, we need an 
efficient algorithm that provides accurate
inference about $\bmmu$ and $\bmpi$. To reach this goal, we 
propose an EM algorithm based on variational Bayesian inference, 
which, through a large body of application studies, is known to 
provide comparably accurate results as other well developed Bayesian 
techniques like Gibbs sampling, but with much faster 
speed~\citep{Logsdon2010avariational, Gazal2012accuracy, 
Tran2017variational, Frazier2022variational}.

In the Bayesian framework, parameter inference is based on the
posterior distributions given the observed
data~\citep{Jordan1999introduction, Westling2019estimation}.
We write out the posterior distribution of the $\mathrm{BM}^2$ 
as follows:
\begin{equation}\label{eq:41}
	p(\bmpi, \bmZ \given \bmmu, \bmR, \bmTheta) = \frac{p(\bmpi,
		\bmZ, \bmR \given \bmmu, \bmTheta)}{\int \sum_{\bmZ} 
		p(\bmpi,
		\bmZ \given \bmmu, \bmR, \bmTheta) \, {\rm d}\bmpi},
\end{equation}
where the denominator is obtained by summing over all possible
membership indicators $\bmZ$, and followed by the integration 
over
all possible $\bmpi$. Unfortunately, the closed form of
Equation~\eqref{eq:41} is unavailable, and its computational
complexity is given by $O(N^{2} M^{2} (K-1)^{2})$. We circumvent 
the
computational challenges by proposing an efficient and scalable
variational EM algorithm, to 
approximate the posterior distributions of $\bmpi$ and $\bmZ$, and
simultaneously to estimate the block level array $\bmmu$.

\subsection{Variational E-step}
\label{sec:Estep}

The fundamental idea of variational methods is to posit a
distribution of the variables of interest (i.e., $\bmpi$ and 
$\bmZ$)
with a set of free variational parameters, and then to choose the
parameters such that the {\em Kullback-Leibler} (K-L) divergence
between the approximation distribution and the true posterior
distribution is minimized. Specifically, let $\{\bmgamma_{i}^{U},
\bmgamma_{j}^{I}, \bmphi_{\itoj}^{U}, \bmphi_{\ifromj}^{I}\}$ be 
a
set of free variational parameters. We denote the variational
distribution family by $\mathcal{Q}$. We specify the variational
distribution $q \in \mathcal{Q}$ via a standand mean-field
variational method; that is
\begin{align}
	q(\bmpi, \bmZ) &= q(\bmpi, \bmZ \given \bmgamma_i^{U},
	\bmgamma_{j}^{I}, \bmphi_{\itoj}^{U}, \bmphi_{\ifromj}^{I})
	\nonumber
	\\ \label{eq:42} &= \prod_{i = 1}^{N}
	q(\bmpi_i^{U} \given \bmgamma_{i}^{U}) \prod_{j = 1}^{M}
	q(\bmpi_j^{I} \given \bmgamma_{j}^{I}) \prod_{R_{ij} \in 
		\bmR}
	q(\bmZ_{\itoj}^U \given \bmphi_{\itoj}^{U})q(\bmZ_{\ifromj}^I
	\given \bmphi_{\ifromj}^{I}) .
\end{align}
In the literature, the distribution $q \in \mathcal{Q}$ (c.f.\
Equation~\eqref{eq:42}) is also known as {\em factorized
	approximation}~\citep{Bishop2006pattern}. As mentioned, our 
goal to
find the best variational distribution $q^* \in \mathcal{Q}$ such
that K-L divergence between $q^*$ and the true posterior
distribution is minimized given the block level array $\bmmu$, 
i.e.,
\[q^* = {\arg \min}_{q \in \mathcal{Q}} {\rm KL}(q(\bmpi, \bmZ)
\, \| \, p(\bmpi, \bmZ \given \bmR, \bmmu, \bmTheta)),\]
where
\[{\rm KL}(q(\bmpi, \bmZ) \, \| \, p(\bmpi, \bmZ \given \bmR, 
\bmmu,
\bmTheta)) = \int_{\bmpi} \sum_{\bmZ} q(\bmpi, \bmZ) \log
\frac{q(\bmpi, \bmZ)}{p(\bmpi, \bmZ \given \bmR, \bmmu, 
	\bmTheta)}
\, {\rm d}\bmpi.\]

It is known that the K-L divergence is minimized if and only if
the distribution $q$ conditional on variational parameters is
identical to $p(\bmpi, \bmZ \given \bmR, \bmmu, \bmTheta)$ if 
there
is no constraint imposed on the variational family $\mathcal{Q}$.
However, this is impractical as the true posterior distribution
itself is intractable. Alternatively, we introduce a practical
procedure for approximating $q$, introduced by
\citet{Latouche2012variational} and \citet{Blei2017variational}.

According to the justifications from~\citet[Section
10]{Bishop2006pattern}, the log-likelihood of $\mathrm{BM}^2$ 
(c.f.\
Equation~\eqref{eq:model}) can be expressed as the composition of
the {\em evidence lower bound}
(ELBO) and K-L divergence:
\[\log p(\bmR, \bmpi, \bmZ \given \bmTheta, \bmmu) =
\mathcal{L}(q(\bmpi, \bmZ); \bmmu, \bmTheta) + {\rm KL}(q(\bmpi,
\bmZ) \, \| \, p(\bmpi, \bmZ \given
\bmR, \bmmu, \bmTheta)),\]
where the ELBO is defined as
\begin{equation}\label{eq:43}
	\mathcal{L}(q(\bmpi, \bmZ); \bmmu, \bmTheta) = \int_{\bmpi}
	\sum_{\bmZ} q(\bmpi, \bmZ) \log \frac{p(\bmpi, \bmZ, \bmR 
		\given
		\bmmu, \bmTheta)}{q(\bmpi, \bmZ)} \, {\rm d}\bmpi.
\end{equation}
Minimizing the K-L divergence is therefore equivalent to 
maximizing
the ELBO. More precisely, in the variational E-step, we maximize 
the
ELBO in Equation~\eqref{eq:43} while assuming that the block 
level
array to be fixed. As such, the inference of $\bmpi$ and
$\bmZ$ has become an optimization problem. The next two 
propositions
are obtained by directly applying the mean-field
theory~\citep{Jordan1999introduction}, for which the proofs are
respectively given in~\ref{app:prop1} and~\ref{app:prop2}.

\begin{proposition}
	\label{prop:1}
	For each user $i$, let $U_i := \{j \given R_{ij} \in \bmR\}$ 
	and
	$I_j := \{j \given R_{ij} \in \bmR\}$ be the set of items 
	rated
	by $i$ and the set of subjects rating item $j$, respectively.
	Given $\bmTheta$, the variational distributions (of $\bmpi$)
	that maximize the ELBO of the $\mathrm{BM}^2$ model are 
	given by
	\begin{align*}
		q(\bmpi_{i}^{U} \given \bmgamma_{i}^{U}) &= {\rm
			Dirichlet}(\bmpi_{i}^{U} ; \bmgamma_{i}^{U}), \qquad 
		i = 1, 2,
		\ldots, N;\\
		q(\bmpi_{i}^{I} \given \bmgamma_{i}^{I}) &= {\rm
			Dirichlet}(\bmpi_{j}^{I} ; \bmgamma_{j}^{I}), \qquad 
		j = 1, 2,
		\ldots, M,
	\end{align*}
	where $\bmgamma^U_{i} := (\gamma^{U}_{ik})_{k = 1}^{K}$ and
	$\bmgamma^I_{j} := (\gamma^{I}_{jl})_{l = 1}^{L}$ with
	\[\gamma_{ik}^{U} = \alpha_{k} + \sum_{j \in U_{i}}
	\phi_{\itoj, k}^{U} \qquad \textrm{and} \qquad 
	\gamma_{jl}^{I} =
	\beta_{l} + \sum_{i \in I_{j}}\phi_{\ifromj, l}^{I}.\]
\end{proposition}

Proposition~\ref{prop:1} gives the explicit distributions of
$\bmpi_i^{U}$ (conditional on $\bmgamma_i^U$) and $\bmpi_j^{I}$
(conditional on $\bmgamma_j^{I}$) for updating $q(\bmpi_{i}^{U}
\given \bmgamma_{i}^{U})$ and $q(\bmpi_{i}^{I} \given
\bmgamma_{i}^{I})$ in Equation~\eqref{eq:42}. Moreover,
$\bmgamma_i^U$ and $\bmgamma_j^I$ are updated through their
respective expressions in Proposition~\ref{prop:1} at each
iteration. In the next proposition, we derive the 
variational
distributions for the remaining two terms in Equation~\eqref{eq:42} 
as
well as the expressions of $\bmphi_{\itoj}^{U}$ and
$\bmphi_{\ifromj}^{I}$.

\begin{proposition}
	\label{prop:2}
	Let $\psi(\cdot)$ be the derivative of the standard log-gamma
	function given by $\psi(x) = \frac{\mathrm{d}}{\mathrm{d} x}
	\log{\Gamma(x)}$. Conditional on the block level array 
	$\bmmu$,
	variational parameters
	$\bmgamma_i, i = 1, 2, \ldots, N$ and $\bmgamma_{j}, j = 1, 
	2,
	\ldots, M$, and rating matrix $\bmR$, the variational
	distributions of ($\bmZ$) that maximize the	ELBO of the 
	$\mathrm{BM}^2$
	model are given by
	\begin{align*}
		q(\bmZ_{\itoj}^{U} \given \bmphi_{\itoj}^{U}) &= {\rm
			Multinomial}(\bmZ_{\itoj}^{U} ; \bmphi_{\itoj}^{U}),
		\qquad i = 1, 2, \ldots, N;\\
		q(\bmZ_{\ifromj}^{I} \given \bmphi_{\ifromj}^{I}) &= {\rm
			Multinomial}(\bmZ_{\ifromj}^{I} ; 
		\bmphi_{\ifromj}^{I}),
		\qquad j = 1, 2, \ldots, M;
	\end{align*}
	where $\bmphi^U_{\itoj} := (\phi^{U}_{\itoj, k})_{k = 1}^{K}$
	and $\bmphi^I_{\ifromj} := (\phi^{I}_{\ifromj,l})_{l = 
		1}^{L}$
	with
	\begin{align*}
		\phi^{U}_{\itoj, k} &\propto
		\exp\left\{\psi(\gamma_{ik}^{U}) - \psi\left(\sum_{k =
			1}^{K} \gamma_{ik}^{U}\right) + \sum_{l = 1}^{L}
		\sum_{s = 1}^{S} R_{ij} \phi_{\ifromj, l}^{I} 
		\log{\mu_{kl,
				s}}\right\} \\
		\phi^{I}_{\ifromj, k} &\propto
		\exp\left\{\psi(\gamma_{jl}^{I}) - \psi\left(\sum_{l =
			1}^{L} \gamma_{jl}^{I}\right) + \sum_{k = 1}^{K}
		\sum_{s = 1}^{S} R_{ij} \phi_{\itoj, k}^{U} \log{\mu_{kl,
				s}}\right\}
	\end{align*}
\end{proposition}
We keep the block structure $\bmmu$ fixed throughout the 
variational
E-step, but update the variational parameters 
$\{\bmgamma_{i}^{U},
\bmgamma_{j}^{I}, \bmphi_{\itoj}^{U}, \bmphi_{\ifromj}^{I}\}$
iteratively pertaining to the relations developed in
Propositions~\ref{prop:1} and~\ref{prop:2}. 

\subsection{Variational M-step}
\label{sec:Mstep}

We have derived the variational distributions of all components 
of
$q(\bmpi, \bmZ)$ (c.f.\ Equation~\eqref{eq:42}). To maximize the
ELBO defined in Equation~\eqref{eq:43}, we start with giving its
explicit expression, as shown in Proposition~\ref{prop:3} below. 
Prior 
to
presenting Proposition~\ref{prop:3}, we introduce two utility
functions, each of which plays a very important role therein. 
Given
a $D$-dimensional vector $\bmx = (x_1, x_2, \ldots, x_D)$, we 
define
\begin{align*}
	f_1(\bmx) &= \log \Gamma\left(\sum_{d = 1}^{D} x_d\right) -
	\sum_{d = 1}^{D} \log \Gamma(x_d),
	\\ f_2(x_d, \bmx) &= \psi(x_d) - \psi\left(\sum_{d = 1}^{D}
	x_d\right).
\end{align*}
Function $f_{1}(\bmx)$ can be regarded as the logarithm of the
constant term of a Dirichlet distribution of dimension $D$ (i.e.,
$\bmx$), whereas function $f_{2}(x_d, \bmx)$ is interpreted as 
the
marginal expectation of $x_d$. We are now ready to present the
proposition, the proof of which is given in~\ref{app:prop3}.
\begin{proposition}
	\label{prop:3}
	Given the block level structure $\bmmu$ and the variational
	parameters $\{\bmgamma, \bmphi\} := \{\bmgamma_{i}^{U},
	\bmgamma_{j}^{I}, \bmphi_{\itoj}^{U},
	\bmphi_{\ifromj}^{I}\}$, the ELBO of the $\mathrm{BM}^2$, 
	comprised of
	four components, is given by
	\[\mathcal{L}(q(\bmpi, \bmZ); \bmmu, \bmTheta) =
	\mathcal{L}_{c} + \mathcal{L}_{\bmgamma} + 
	\mathcal{L}_{\bmphi,
		\bmgamma} + \mathcal{L}_{\bmphi, \bmmu},\]
	where
	\begin{align*}
		\mathcal{L}_{c} &= N f_1(\bmalpha) + M f_1(\bmbeta),
		\\ \mathcal{L}_{\bmgamma} &= \sum_{i = 1}^{N} 
		\left(\sum_{k
			= 1}^{K} \left(\alpha_k - \gamma_{ik}^{U}\right)
		f_2\left(\gamma_{ik}^{U}, \bmgamma_{i}^{U}\right)\right) 
		+
		\sum_{j = 1}^{M} \left(\sum_{l = 1}^{L} \left(\beta_l -
		\gamma_{jl}^{I}\right) f_2\left(\gamma_{jl}^{I},
		\bmgamma_{j}^{I}\right)\right),
		\\ \mathcal{L}_{\bmphi, \bmgamma} &= \sum_{i,j}
		\left(\sum_{k = 1}^{K} \phi_{\itoj, k}^{U}
		\left(f_2\left(\gamma_{ik}^{U}, \bmgamma_{i}^{U}\right) -
		\log \phi_{\itoj, k}^{U}\right) + \sum_{l = 1}^{L}
		\phi_{\ifromj, l}^{I} \left(f_2\left(\gamma_{jl}^{I},
		\bmgamma_{j}^{I}\right) -
		\log \phi_{\ifromj, l}^{I}\right)\right),
		\\ \mathcal{L}_{\bmphi, \bmmu} &= \sum_{R_{ij} \in \bmR}
		\sum_{k = 1}^{K} \sum_{l = 1}^{L} \sum_{s = 1}^{S}
		\phi_{\itoj, k}^{U} \phi_{\ifromj, l}^{I} c_{ij,s} \log
		\mu_{kl, s}.
	\end{align*}
\end{proposition}
As shown in Proposition~\ref{prop:3}, we divide the ELBO defined
in Equation~\eqref{eq:43} into four parts, where 
$\mathcal{L}_{c}$
is a constant only related to the hyperparameters,
$\mathcal{L}_{\bmgamma}$ depends on the variational parameters
$\bmgamma_i^{U}$ and $\bmgamma_j^{I}$ only, and
$\mathcal{L}_{\bmphi, \bmgamma}$ relies on all hyperparameters, 
and
$\mathcal{L}_{\bmphi, \bmmu}$ is associated with both
hyperparameters as well as block structure. 
Proposition~\ref{prop:3}
provides a tractable alternative to the estimation of posterior
distribution in Equation~\eqref{eq:41}. The convergence rate of 
the
proposed variational EM algorithm is determined via the 
formulation
of ELBO.

Lastly, we provide a scheme updating $\bmmu$ at each
iteration to complete the variational M-step.

\begin{proposition}
	\label{prop:4}
	Given the variational parameters $\{\bmgamma, \bmphi\}$ and 
	the
	rating matrix $\bmR$, we maximize the defined ELBO by 
	updating
	$\bmmu$ iteratively with
	\[
	\mu_{kl, s} = \frac{\sum_{i, j} \phi_{\itoj, k}^{U}
		\phi_{\ifromj, l}^{I} R_{ij} \bm{1}(R_{ij} = 
		s)}{\sum_{i, j}
		\phi_{\itoj, k}^{U} \phi_{\ifromj, l}^{I}},
	\]
	for $k = 1, 2, \ldots ,K$, $l = 1,2, \ldots, L$ and $s = 1, 
	2,
	\ldots, S$.
\end{proposition}

See~\ref{app:prop4} for the proof of 
Proposition~\ref{prop:4}. The interpretation of the proposed 
estimator for $\bmmu$ is straightforward, as the variational 
parameter $\phi_{\itoj, k}^{U}$
approximates the probability that user $i$ belongs to
cluster $k$ when he/she rates the item $j$, and $\phi_{\ifromj, 
	l}^{I}$
is interpreted analogously. In what follows, the denominator
integrates all the possibilities that users are from cluster $k$ 
and
items are categorized to cluster $l$ regardless of the ratings,
whereas the numerator accounts for an additional factor that the
rating score is equal to some given $s$. While updating the block
structure $\bmmu$ (pertaining to Proposition~\ref{prop:4}), we 
keep
the current estimates of the variational parameters unchanged, 
such
that all the parameters which contribute to the ELBO are updated
iteratively and exchangeably until convergence. 

\subsection{Variational EM algorithm}
\label{sec：VEM}

Based on Propositions \ref{prop:1} through \ref{prop:4}, we are
ready to propose our variational EM algorithm, the pseudo
codes of which are given in Algorithm \ref{alg:VEM}. The 
convergence
of the proposed algorithm is naturally guaranteed by the process 
of
variational inference~\citep{Blei2017variational}. Starting from 
the
initial values of the variational parameters, $\bmgamma^{(0)} :=
\{\bmgamma_{i}^{U, (0)}, \bmgamma_{j}^{I, (0)}\}$ and 
$\bmphi^{(0)}
:= \{\bmphi_{\itoj}^{U, (0)}, \bmphi_{\ifromj}^{I, (0)}\}$, and 
the
initial block structure $\bmmu^{(0)}$, we repeat the following 
two
procedures until reaching convergence:
\begin{enumerate}
	\item Update $\bmphi$ based on the given observed ratings 
	$\bmR$;
	\item Update $\bmgamma$ and $\bmmu$ based on the current
	estimates of $\bmphi$, ratings $\bmR$ and hyperparameters
	$\bmalpha$ and $\bmbeta$.
\end{enumerate}
\begin{algorithm}[tbp]
	\caption{Pseudo codes for the variational EM algorithm for
		$\mathrm{BM}^2$.}
	\label{alg:VEM}
	\KwIn{Rating matrix $\bmR$, hyperparameters $\bmTheta =
		\{\bmalpha, \bmbeta\}$.}
	\KwOut{Block structure $\bmmu$, variational parameters
		$\bmgamma$ and $\bmphi$.}
	Initialize $\bmmu^{(0)}$, $\bmgamma^{(0)}$, $\bmphi^{(0)}$\;
	\While{$t > 0$}{
		\For{$k = 1$ \KwTo $K$}{
			Update $\bmphi_{\itoj}^{U, (t + 1)} \propto
			g(\bmgamma_{i}^{U, (t)},
			R_{ij}, \bmphi_{\ifromj}^{I, (t)}, \bmmu^{(t)})$}
		Normalize $\bmphi_{\itoj}^{U, (t + 1)}$\;
		\For{$l = 1$ \KwTo $L$}{
			Update $\bmphi_{\ifromj}^{I, (t + 1)} \propto
			h(\bmgamma_{j}^{I, (t)},
			R_{ij}, \bmphi_{\itoj}^{U, (t + 1)}, \bmmu^{(t)})$}
		Normalize $\bmphi_{\ifromj}^{I, (t + 1)}$\;
		\For{$i = 1$ \KwTo $N$}{
			Update $\bmgamma_{i}^{U, (t)} = \bmalpha + \sum_{j 
				\in
				U_{i}} \bmphi_{\itoj}^{U, (t)}$\;	
		}
		\For{$j = 1$ \KwTo $M$}{
			Update $\bmgamma_{j}^{I, (t)} = \bmbeta + \sum_{i \in
				I_{j}} \bmphi_{\ifromj}^{I, (t)}$\;	
		}
		\For{$s = 1$ \KwTo $S$}{
			Update $\mu_{kl, s}^{(t)} = \frac{\sum_{i, j}
				\phi_{\itoj, k}^{U, (t)} \phi_{\ifromj, l}^{I, 
					(t)}
				R_{ij} \bm{1}(R_{ij} = s)}{\sum_{i, j} 
				\phi_{\itoj,
					k}^{U, (t)} \phi_{\ifromj, l}^{I, (t)}}$ for 
			all $k = 1,
			2, \ldots K$ and $l = 1, 2, \ldots, L$\;	
		}
		$t \leftarrow t - 1$\;
	}		
\end{algorithm}

From Proposition \ref{prop:2}, we notice that the variational
parameters $\bmphi_{\itoj}^{U}$ and $\bmphi_{\ifromj}^{I}$ are
dependent upon each other. Besides, we need the knowledge of 
$\bmphi_{\itoj}^{U}$ and $\bmphi_{\ifromj}^{I}$ for updating the 
other variational parameters as well as the block structure 
according to Propositions~\ref{prop:1} and~\ref{prop:4}. Hence, 
in 
Algorithm~\ref{alg:VEM}, we generate the next iterations for 
$\bmphi_{\itoj}^{U}$ and $\bmphi_{\ifromj}^{I}$ prior to the 
rest. 
Specifically, the functions $g$ (at step 4) and $h$ (at step 7) 
correspond to those expressed in Proposition~\ref{prop:2}, 
respectively. We note that the updated $\bmphi_{\itoj}^{U}$ and 
$\bmphi_{\ifromj}^{I}$ need to be normalized at the end of each 
iteration in order to meet their distributional properties.

The computational complexity of the proposed 
algorithm
is $O(N + M + (S + R_0)(K + L))$, where the $R_0$ represents the
actual number of observed ratings. In practice, $R_0$ is usually
much larger than $N$ or $M$, whereas $K$ and $L$ are relatively
small constants, so that the proposed algorithm is
linearly scalable with respect to $R_0$.

\subsection{Rating prediction}
\label{sec:rating}

Our primary interest is to conduct rating prediction, which is based 
on the mixed-membership vector $\bmpi$ and block structure $\bmmu$. 
We obtain the estimates of $\bmpi$ from the variational 
distributions.
They are respectively given by
\begin{align*}
	\hat{\pi}_{ik}^{U} = \frac{\gamma_{ik}^{U}}{\sum_{k = 1}^{K} 
		\gamma_{ik}^{U}}, i = 1, 2, \ldots,N; k = 1, 2, \ldots 
	,K,
	\\ \hat{\pi}_{jl}^{I} = \frac{\gamma_{jl}^{I}}{\sum_{l = 
			1}^{L}\gamma_{jl}^{I}}, j = 1, 2, \ldots ,M; l = 1, 
	2, \ldots, L.
\end{align*}

Following the idea that the probability of each rate is governed 
by the clusters that the user and the item respectively belong 
to~\citep{Godoy2016accurate}, we are 
able to predict the rating of item $j$ by user $i$ as follows
\[
\hat{R}_{ij} = \argmax_{C_s \in \mathcal{C}}p(R_{ij} = C_s) = 
\argmax_s \sum_{k = 1}^{K} \sum_{l = 1}^{L} \hat{\pi}_{ik}^{U} 
\hat{\mu}_{kl, s} \hat{\pi}_{jl}^{I},
\]
which is the rate associated with the highest block-wise 
probability. It is worth to mention that $\hat{R}_{ij}$ can be
viewed as a maximum a posterior (MAP) estimate in the context of 
Bayesian theory. As the proposed model allows for more than one
memberships (i.e., soft clustering), it suffices to report the
mixed membership vector which contains more information, but a
single-value estimate (i.e., $\hat{R}_{ij}$) is usually used for
prediction accuracy assessment. 

\section{Simulations}
\label{sec:sim}

In this section, we assess the performance of the proposed 
$\mathrm{BM}^2$ based on prediction accuracy via extensive 
simulations. 
In particular, we have included some outliers in the simulated 
data to show the robustness of our algorithm. 

The data generation procedure is done under the fundamental 
principle of collaborative filtering that similar users tend to 
give similar ratings to similar items, where the similarities of 
users and items are reflected in their respective clusters. For 
instance, tough raters are likely to give lower rates to an item 
than benevolent raters; also, raters usually give different rates to 
items of different qualities. More specifically, we 
consider $N = 300$ users and $M = 200$ items, altogether leading 
to $500$ nodes in the associated bipartite network. For 
simplicity, we assume that the number of user cluster $K$ equals 
the number of item clusters $L$ in 
the present simulation study. 
We have considered $K = 
L \in \{5, 7, 9\}$, where the three scenarios respectively 
correspond to standard, extensive and extremely extensive rating 
systems. For instance, for $K = L = 5$, each user falls into one 
of the following groups: very strict, strict, modest, generous, 
very generous, whereas each item falls into one of the following 
categories: very dissatisfied, dissatisfied, neutral, satisfied, 
very satisfied.

Different from classical MMSBM which requires one block matrix 
representing network community 
structure~\citep{Airoldi2008mixed, 
	Karrer2011stochastic}, the proposed $\mathrm{BM}^2$ model 
requires 
the 
generation of $S$ block matrices, $\bmmu := \{\bmmu_{1}, 
\bmmu_{2}, 
\ldots,\bmmu_{S}\}$, each of which corresponds a specific rate. 
Let 
us take a look at $\bmmu_1$ as example:
\[
\bmmu_1 := \begin{pmatrix}
	0.650 & 0.450 & 0.250 & 0.150 & 0.100 \\
	0.450 & 0.250 & 0.050 & 0.050 & 0.050 \\
	0.100 & 0.100 & 0.050 & 0.050 & 0.050 \\
	0.100 & 0.100 & 0.020 & 0.020 & 0.020 \\
	0.100 & 0.100 & 0.020 & 0.020 & 0.020
\end{pmatrix}.
\]
The element $\mu_{kl,1}$ in $\bmmu_1$ is the probability that a 
user 
from cluster $k$ rates $1$ to an item from cluster $l$. For the 
simulations, we assign $\mu_{11, 1} = 0.650$, referring to a 
high 
probability that a ``strict'' user tends to give the lowest rate 
to 
an item from the ``very dissatisfied'' category. Row-wise, this 
probability decreases, or does not increase, as the users in the 
latter categories are generally more forgiving. Column-wise, 
this 
probability does not increase either, because item quality 
improves. 
All the block matrices are generated in the same manner. We 
refer 
the readers to~\ref{app:simmu} for the details as well 
as 
for reproducing the simulation results.

We sample the mixed-membership vectors for each user and each 
item 
respectively from ${\rm Multinomial}(1;\bmalpha)$ and ${\rm 
	Multinomial}(1;\bmbeta)$, where the known hyperparameters 
$\bmalpha$
and $\bmbeta$ contain the prior information. For the case of $K 
= L 
= 5$, we set hyperparameters $\bmalpha := 
(0.10,0.20,0.40,0.20,0.10)$ and $\bmbeta := 
(0.10,0.15,0.45,0.25,0.05)$, suggesting that the majority of 
users 
are neutral, and that almost half of the items are of average 
quality. The rating matrix $\bmR$ is then constructed upon 
$\bmmu$, 
$\bmalpha$ and $\bmbeta$. For any pair of $i, j$, we have
\[
\bmc_{ij} \sim {\rm Multinomial}(1;\bm{\delta}),
\] 
where $\bm{\delta}$ is an $S$-long vector consisting of 
$\delta_s = 
\sum_{k = 1}^{K} \sum_{l = 1}^{L} \bmalpha_k \mu_{kl, s} 
\bmbeta_l$ 
for $s = 1, 2, \ldots, S$.

Specifically, we have added a few outliers in 
the data generation process to check the robustness of the 
proposed 
model in contrast to the competing models. For instance, 
generous 
raters are extremely likely to rate $5$ (the highest score for 
the 
scenario of $K = L = 5$) to high quality items. We select $10\%$ 
of 
those rates completely at random, and set them to $1$. 
Analogously, 
we randomly sample $1$'s (about $10\%$, too) rated by critics to 
low 
quality items, and set them to $5$. We adopt the MMSBM proposed 
by 
\citet{Godoy2016accurate} as the competing model in the 
simulation 
study. The MMSBM used an EM algorithm to predict unobserved 
rates, 
and the method has been applied to a large movie rating data. It 
is 
evident the model outperforms standard collaborative filtering 
methods, like matrix factorization. One of the main tasks is to  
investigate whether or not there is a further increase in prediction 
accuracy via the Bayesian methods (with both non-informative and 
informative priors) compared to the MMSBM in the presence of 
outliers.

\begin{table}[htbp]
	\centering
	\caption{Comparison of MMSBM, $\mathrm{BM}^2$ (with 
		non-informative 
		prior) and $\mathrm{BM}^2$* (with informative prior); 
		the unobserved data 
		proportion is $(1 - \eta) = 0.8$.}\label{tab:1}
	\begin{tabular}{clccc}
		\toprule
		& & \multicolumn{3}{c}{Evaluation criterion} 
		\\
		\cmidrule(lr){3-5} 
		Cluster number & Model & MAE (SE) & MSE (SE) & AR (SE)\\
		\midrule
		\multirow{3}{*}{$K = L = 5$} & MMSBM & $0.7955(0.0199)$ 
		& 
		$1.2972(0.0542)$ &	$\blue{0.4114}(0.0111)$
		\\
		& $\mathrm{BM}^2$ & $0.7994(0.0192)$ & $1.2807(0.0552)$ 
		& 
		$0.4012(0.0100)$
		\\
		& $\mathrm{BM}^2$* & $\blue{0.7940} (0.0193)$ & 
		$\blue{1.2676} 
		(0.0568)$ &	$0.4022(0.0093)$ 
		\\ 
		\midrule
		\multirow{3}{*}{$K = L = 7$} & MMSBM & $0.8071(0.0209)$ 
		& 
		$1.3225(0.0497)$ & $\blue{0.4067}(0.0128)$
		\\
		& $\mathrm{BM}^2$ & $0.8068(0.0195)$ & $1.2889(0.0489)$ 
		& 
		$0.3962(0.0108)$ 
		\\
		& $\mathrm{BM}^2$* & $\blue{0.7983} (0.0179)$ & 
		$\blue{1.2661} 
		(0.0475)$ &	$0.3983(0.0094)$  
		\\ 
		\midrule
		\multirow{3}{*}{$K = L = 9$} & MMSBM & $0.7964(0.0128)$ 
		& 
		$1.2428(0.0335)$ & $0.3947(0.0077)$
		\\
		& $\mathrm{BM}^2$ & $0.7864(0.0125)$ & $1.2018(0.0352)$ 
		& 
		$0.3934(0.0064)$ 
		\\
		& $\mathrm{BM}^2$* & $\blue{0.7727} (0.0122)$ & 
		$\blue{1.1730} 
		(0.0358)$ &	$\blue{0.4003} (0.0057)$  
		\\ 
		\bottomrule
	\end{tabular}
\end{table}

Moreover, for the proposed $\mathrm{BM}^2$, we consider two 
different 
sub-types: namely non-prior $\mathrm{BM}^2$ and correct-prior 
$\mathrm{BM}^2$. For 
the 
former sub-type, we assume a lack of knowledge of true 
$\bmalpha$ 
and $\bmbeta$. Accordingly, for the case of $K = L = 5$, we use 
$(1/5, 1/5, 1/5, 1/5, 1/5)$ as the probability vector for 
sampling 
the groups for the users, and for the items as well. However, 
for 
the latter sub-type, we use the correct $\bmalpha$ and 
$\bmbeta$ as given in the preceding paragraph as the inputs to 
implement the proposed algorithm. We account for these two types 
of 
$\mathrm{BM}^2$ in order to assess the performance of the 
proposed model 
when 
there exists some correct auxiliary information as prior. 

We adopt three criteria for evaluating the prediction accuracy; 
namely, they are mean squared error (MSE), mean absolute error (MAE) 
and 
accuracy rate (AR), which are respectively defined as
\begin{align*}
	{\rm MAE} &= \frac{1}{|\bmR|} \sum_{ij} |\hat{R}_{ij} - R_{ij}|
	\\ {\rm MSE} &= \frac{1}{|\bmR|} \sum_{ij} (\hat{R}_{ij} - 
	R_{ij})^2
	\\ {\rm AR} &= \frac{1}{|\bmR|}\sum_{ij} \bm{1}_{\{\hat{R}_{ij} 
	= 
		R_{ij}\}},
\end{align*}
where $\bm{1}_{\{\cdot\}}$ is the standard indicator function. These 
three criteria are selected since MAE evaluates the closeness 
between predictions and true values, MSE is a quantity measuring the 
squared error loss, and AR intuitively tells the prediction 
accuracy. It is necessary to look into all three criteria together 
for an overall model performance assessment. To compute MAE, MSE and 
AR, we need the true values of $R_{ij}$ as ground truth. In our data 
generation procedure, we generate the ratings for each pair of the 
users and items, but only make a proportion of ratings that are 
observed, governed by a specified parameter $\eta \in (0, 1)$. In 
other words, $(1 - \eta)$ of the ratings are hidden, and will be 
predicted by the proposed model as well as other competing methods. 
The assessment of model performance is based on prediction accuracy 
and estimation variations through MAE, MSE and AR. We would not 
treat these unobserved ratings as missing data in the present study, 
as we view these unobserved data are simply caused by no interaction 
between the users and the items. On the other hand, the model 
performance evaluation is based on the assumption that the users 
were to rate the items that had not yet been rated by them. Thus, 
our focus of model performance evaluation is on unobserved rate 
predictions, but not on the accuracy of variational approximation.

The simulation is based on $100$ independent replicates, and the 
results are presented in Table~\ref{tab:1}. To 
give a comprehensive study, we have also added the simulation 
results for $K = L = 7$ and $K = L = 9$ to Table~\ref{tab:1}, 
where the associated parameter settings can be found 
in~\ref{app:simmu}. We have observed 
obvious improvements from MMSBM to 
$\mathrm{BM}^2$ (without informative prior) based off MAE 
and MSE. Besides, more improvements appear in $\mathrm{BM}^2$* (with 
correct prior). There is no consistent pattern in the measure of AR. 
$\mathrm{BM}^2$ 
outperforms the other two methods ($\mathrm{BM}^2$* and MMSBM) for 
$K = L = 5$ and $K = L = 7$, but MMSBM narrowly beats the rest for 
$K = L = 9$. In general, the 
difference in AR across the three methods is negligible. Thus, based 
on the 
overall performance (according to all three criteria), we conclude 
that the proposed method is preferred 
to MMSBM, especially when there exits correct auxiliary 
information. 

It is well known that variational EM algorithms are 
efficient~\citep{Airoldi2008mixed}. To confirm, we provide the 
running time based on $100$ independent replicates for $K = L = 5$ 
as an example in Table~\ref{tab:time}, where we see that ${\rm 
	BM}^2$* costs least time among all despite that all algorithms 
	are 
quite efficient. It takes a bit longer for ${\rm BM}^2$ to converge 
owing to non-informative prior. We omit the running time tables for 
$K = L = 7$ and $K = L = 9$ since they present similar patterns. It 
is worth mentioning that we recommend using informative prior only 
if some useful knowledge or supporting information is available; 
otherwise, non-informative priors should be considered. 

\begin{table}[ht]
	\centering
	\caption{Running time based on $100$ independent replicates for 
	$K = L = 5$}
	\label{tab:time}
	\begin{tabular}{lccc}
		\toprule
		& \multicolumn{3}{c}{Model} 
		\\
		\cmidrule(lr){2-4} 
		& MMSBM & ${\rm BM}^2$ & ${\rm BM}^2$*\\
		\midrule
		Time (seconds) & $242.94$ & $636.04$ & $90.49$
		\\
		\bottomrule
	\end{tabular}
\end{table}

\begin{figure}[tbp]
	\includegraphics[width = \textwidth]{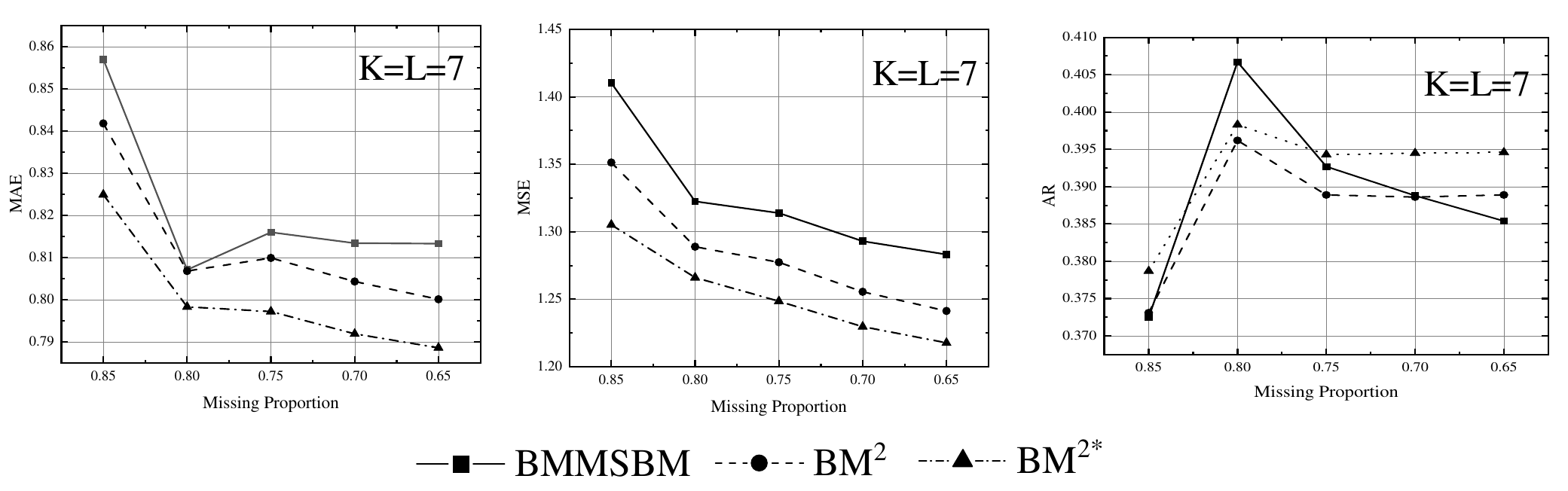}
	\caption{Dynamics of MAE, MSE and AR with respect to 
		different unobserved data proportions.}
	\label{fig:sensitivity}
\end{figure}

In addition, we carry out a sensitivity analysis that assesses 
the performance of the models with different unobserved data
proportions, where the results for $K = L = 7$ are presented in 
Figure~\ref{fig:sensitivity}. From MAE and MSE, we see that 
$\mathrm{BM}^2$ performs consistently better than MMSBM. 
Especially when correct prior information is available, 
$\mathrm{BM}^2$ has the lowest MAE and MSE regardless of unobserved 
data 
proportion. As expected, with the decrease of unobserved data 
proportion, both MAE and MSE values of $\mathrm{BM}^2$ are 
reduced, since more available data help improve the prediction 
accuracy. For AR values, we do not observe significant 
deviations across three models, since the unit of vertical axis 
(for AR) is $0.005$. Although MMSBM has the highest AR value 
when the unobserved data proportion is $0.8$, $\mathrm{BM}^2$ with 
correct prior performs better than the other two models for the 
rest. Even for the case of $0.8$ unobserved data proportion, we 
still 
recommend $\mathrm{BM}^2$ with correct prior information. Its AR 
value is close to that for MMSBM, but lower MAE and MSE jointly 
imply that its error between the predictions and true ratings 
are small.	

Although the determination of cluster number is not our primary 
focus, it is an inevitable procedure prior to implementing the 
proposed algorithm for prediction. We suggest to determine $K$ and 
$L$
via {\em cross-validation} (CV). Specifically, in the simulation 
study,
we consider a 
five-fold CV method based on MAE to 
find optimal $K = L$. That is, we split the data (ratings) 
evenly into five folds with four training folds and one testing 
fold. For each of the cluster number candidate, we train the 
model on the training data, and then use the MAE based on the 
predicted ratings and true values in the testing data as 
selection criterion. We repeat this procedure for $100$ times, 
and summarize the CV results in Table~\ref{tab:cv}. For each 
considered scenario, the proposed method tends to slightly 
over-estimate 
the number of clusters, which is a typical behavior of CV-based 
approaches~\citep{Gabriel2002le, Fu2020estimating}. Nevertheless, the
selected models never underestimate the number of clusters and 
exhibit comparable predictive performance. As CV-based approach 
focuses on prediction, the most frequently selected model indeed 
leads to the lowest MAE in each setting.

\begin{table}[tbp]
	\label{tab:cv}
	\caption{Five-fold cross-validation results of cluster 
		number selection. Selection frequencies are out of $100$ 
		attempts.}
	\centering
	\footnotesize \setlength{\tabcolsep}{2pt}
	\begin{tabular}{c ccccc | ccccc | ccccc}
		\toprule
		True number & \multicolumn{5}{c}{$K = L = 5$} & 
		\multicolumn{5}{c}{$K = L = 7$} & \multicolumn{5}{c}{$K 
			= L = 9$} 
		\\
		\cmidrule(lr){2-6} \cmidrule(lr){7-11}  
		\cmidrule(lr){12-16}
		Selection & 3 & 4 & 5 & 6 & 7 & 5 & 6 & 7 & 8 & 9 & 
		7 & 8 & 9 & 10 & 11 \\ 
		\midrule
		Frequency & 0 & 0 & 1 & 13 & 86 & 0 & 0 & 4 
		& 15 & 81 & 0 & 0 & 5 & 25 & 70 \\
		MAE mean & 0.62 & 0.59 & 0.57 & 0.56 & 0.55 & 0.60 & 
		0.59 & 0.58 & 0.57 & 0.56 & 0.45 & 0.44 & 0.43 & 0.42 & 
		0.41 \\
		\bottomrule 
	\end{tabular}
\end{table}

\section{MovieLens Data Application}
\label{sec:movelens}

We apply the proposed $\mathrm{BM}^2$ to the MovieLens data, and 
compare 
the 
results with those from MMSBM and three traditional 
collaborative 
filtering algorithms: user-based method, item-based method, and 
probability matrix factorization (PMF). The MovieLens dataset 
belongs to the GroupLens Research Project which aims to 
recommend 
movies via traditional collaborative filtering methods, and it 
is publicly available at 
\url{https://grouplens.org/datasets/movielens}. The data has 
collected the movie ratings from September 19, 1997 to April 22, 
1998, and is maintained by the members of GroupLens lab 
based at the University of Minnesota 
(\url{https://movielens.org/}). The entire MovieLens dataset 
contains more than $20$ million ratings. The MovieLens dataset used 
in the present study is also known as MovieLens $100$K, which 
consists of $943$ users and $1,682$ 
movies, 
leading to a total of $2,625$ nodes in its associated bipartite 
network. The dataset has recorded $10,000$ ratings, 
where each of the users has rated at least $20$ movies based on 
a 
one-to-five scale. Each record in the dataset contains four 
variables: user ID, item ID, rate and timestamp.

To compare the proposed algorithm with the competing methods, we 
report the MAE, MSE and AR for each method, alongside with their 
respective standard errors. Similar to the simulation setup in 
the preceding section, we randomly select $20\%$ of the MovieLens 
dataset to 
form the training set, where the rest are assumed to be unobserved. 
In 
Table~\ref{tab:movielens}, we present the results for four 
traditional collaborative filtering methods, in addition to 
those 
for MMSBM and the proposed $\mathrm{BM}^2$. The first 
collaborative 
filtering 
method that we consider is a naive approach, which uses the 
average 
of an user's all observed ratings to predict all of her 
unobserved 
ratings. The method is popular in the old times owing to its 
simplicity, but usually fails to provide accurate predictions.
User-based~\citep{Lee2006improved} and 
item-based~\citep{Deshpande2004item} models are two other 
classical 
collaborative filtering approaches. The prediction of missing 
rates 
is analogous to nearest neighbor imputation, where the 
similarity 
among the users or items is usually measured via some 
quantitative 
metrics. Specifically, we have adopted cosine 
similarity~\citep{Billsus2000user}, which is related to 
Pearson correlation, in the present analysis. The implementation 
of 
these two methods is programmed in $\mathtt{Matlab}$, the codes of 
which are
available in 
the online supplements. From Table~\ref{tab:movielens}, we see 
that 
there is no significant difference in either MAE or MSE 
between the item- and user-based methods, though both outperform 
the 
naive approach. PMF~\citep{Mnih2007probabilistic, 
	Koren2009matrix} 
is a model-based method decomposing the observed rating matrix 
into 
a user factor matrix and an item factor matrix. Specifically, we 
have applied a coordinate gradient descent method to update the 
(low 
rank) implicit user and item factor matrices iteratively by 
minimizing square loss. The resulting user and item factor 
matrices 
are used to predict the unobserved rates; see the 
$\mathtt{Matlab}$ 
codes in the supplements for details. According to the MAE and 
MSE 
of PMF in Table~\ref{tab:movielens}, the performance of PMF is 
not 
as good as that of either item- or user-based method. Lastly, we 
look into the performance of MMSBM and $\mathrm{BM}^2$ without 
informative 
prior. In addition to MAE and MSE, we are able to report AR for 
these two methods since the predicted rates are integer-valued. 
Both of the user cluster number $K$ and item cluster number $L$ 
are set to $10$ upon the suggestion by~\citet{Godoy2016accurate}. 
$\mathrm{BM}^2$ has a 
smaller MSE value than MMSBM, and they both perform better than 
PMF and the naive approach. Besides, $\mathrm{BM}^2$ has the 
smallest MAE value among all the considered methods, whereas 
their difference in AR is extremely small, rendering 
$\mathrm{BM}^2$ a preferred method for this 
application.   

\begin{table}[tbp]
	\centering
	\caption{Performance of different models with applications 
		to the MovieLens dataset.}\label{tab:movielens}
	\begin{tabular}{lccc}
		\toprule
		& \multicolumn{3}{c}{Evaluation criterion} 
		\\
		\cmidrule(lr){2-4} 
		Model & MAE (SE) & MSE (SE) & AR (SE)\\
		\midrule
		Naive & $1.3269 (0.0014)$ & 
		$2.3271 (0.0049)$ &
		\\
		Item-based & $0.8068 (0.0001)$ & $1.0278 (0.0006)$ & 
		\\
		User-based & $0.8110 (0.0003)$ & $1.0344 (0.0009)$ &
		\\
		PMF & $0.8493 (0.0117)$ & $1.2557 (0.0473)$ &
		\\
		MMSBM & $0.7439 (0.0015)$ & $1.1943 (0.0033)$ & 
		$\blue{0.4352} (0.0009)$
		\\
		$\mathrm{BM}^2$ & $\blue{0.7300} (0.0025)$ & 
		$\blue{1.1613} 
		(0.0075)$ 
		& $0.4417 (0.0011)$
		\\
		\bottomrule
	\end{tabular}
\end{table}

\begin{table}[tbp]
	\centering
	\caption{Clustering summary for the MovieLens 
		dataset via $\mathrm{BM}^2$ 
		model.}\label{tab:user_item_cluster}
	\begin{tabular}{l ccccccccccc}
		\toprule
		User Cluster & $U_1$ & $U_2$ & $U_3$ & $U_4$ & $U_5$ & 
		$U_6$ & $U_7$ & $U_8$ & $U_9$ & $U_{10}$ \\
		\midrule
		Size & 102 & 56 & 131 & 81 & 141 & 107 & 57 & 69 & 111 & 
		88 \\
		Average Rating & 3.52 & 3.79 & 3.58 & 2.81 & 3.97 & 3.29 
		& 3.82 & 3.07 & 3.81 & 3.58 \\
		\midrule 
		\midrule
		Item Cluster & $I_1$ & $I_2$ & $I_3$ & $I_4$ & $I_5$ & 
		$I_6$ & $I_7$ & $I_8$ & $I_9$ & $I_{10}$ \\
		\midrule
		Size & 177 & 128 & 137 & 153 & 166 & 120 & 114 & 137     
		& 86 & 467 \\
		Average Rating & 4.12 & 3.53 & 2.97 & 3.22 & 3.18 & 3.15 
		& 3.68 & 3.86 & 3.80 & 2.60 \\
		\bottomrule
	\end{tabular}
\end{table}

By applying the $\mathrm{BM}^2$ model, each user (and item) is 
associated with a membership vector. We conduct hard clustering 
by assigning cluster membership (to the users and items) 
according to the largest component in the membership vector. We 
present the cluster sizes and the average rating for each 
cluster in Table~\ref{tab:user_item_cluster}. User-wise, we find 
that the lowest average rating is $2.81$ (belonging to $U_4$), 
whereas the highest average rating is $3.97$ (belonging to 
$U_5$). This implies that the users from $U_4$ are likely to be 
very strict, but those from $U_5$ appear to be more generous 
than the rest. On the other hand, the average rating of $I_1$ is 
$4.12$, which much larger than that of $I_{10}$, given by 
$2.60$, suggesting high quality of the movies from $I_1$, but 
low quality of the movies from $I_{10}$.

\begin{figure}[tbp]
	\centering
	\includegraphics[width = 0.85\textwidth]{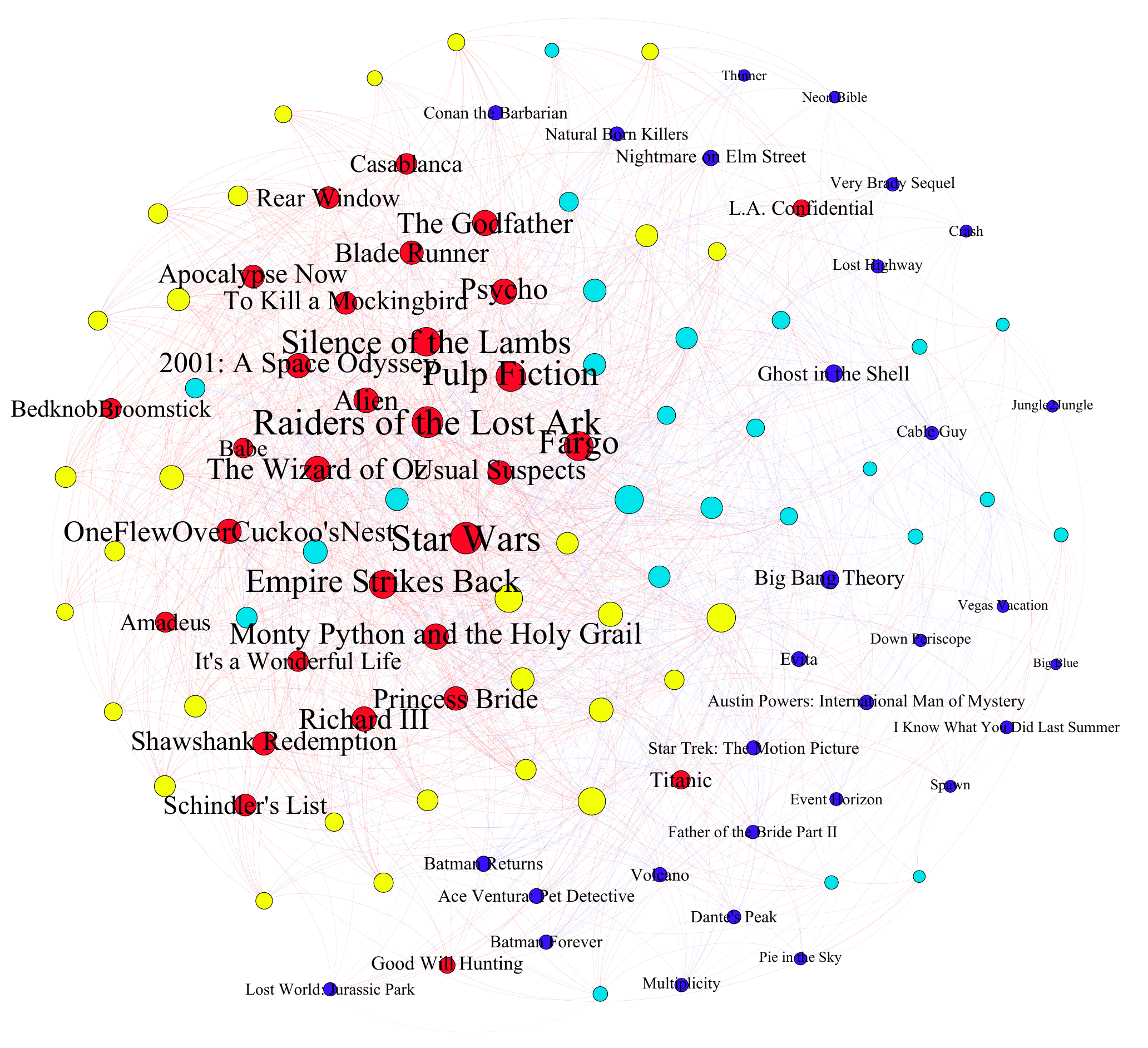}
	\caption{The network based on the users in cluster $U_4$ 
		(cyan) and $U_5$ (yellow) and the movies in cluster $I_1$ 
		(red) and $I_{10}$ (blue). Red and blue edges represent the 
		high ($4$ and $5$) and low ($1$, $2$ and $3$) rates, 
		respectively. Node size is proportional to node degree.} 
	\label{fig:movielens_net}
	4 	\end{figure}

In Figure~\ref{fig:movielens_net}, we give a sub-network 
structure based on the users from $U_4$ and $U_5$ as well as the 
movies from $I_1$ and $I_{10}$. The edges are colored with red 
and blue, respectively representing high ($4$ and $5$) and low 
($1$, $2$ and $3$) rates. The size of the nodes are 
proportional to their degrees, indicating the number of rates 
that the movies have received. We see many movies 
that have high reputations, such as Star Wars, Raiders of the 
Lost Ark, Fargo, Pulp Fiction and Silence of the Lambs, belong 
to $I_{1}$. These high-quality movies are attractive, so tentatively 
will get many good ratings, i.e., attached by red edges. 
However, those movies belonging to 	$I_1$ do not even receive 
too many rates due to lack of attention among the users. In 
average, the sizes of yellow nodes (for $U_5$) are greater than 
those of cyan nodes (for $U_4$) as expected. 

\begin{figure}[tbp]
	\centering
	\includegraphics[width = 0.6\textwidth]{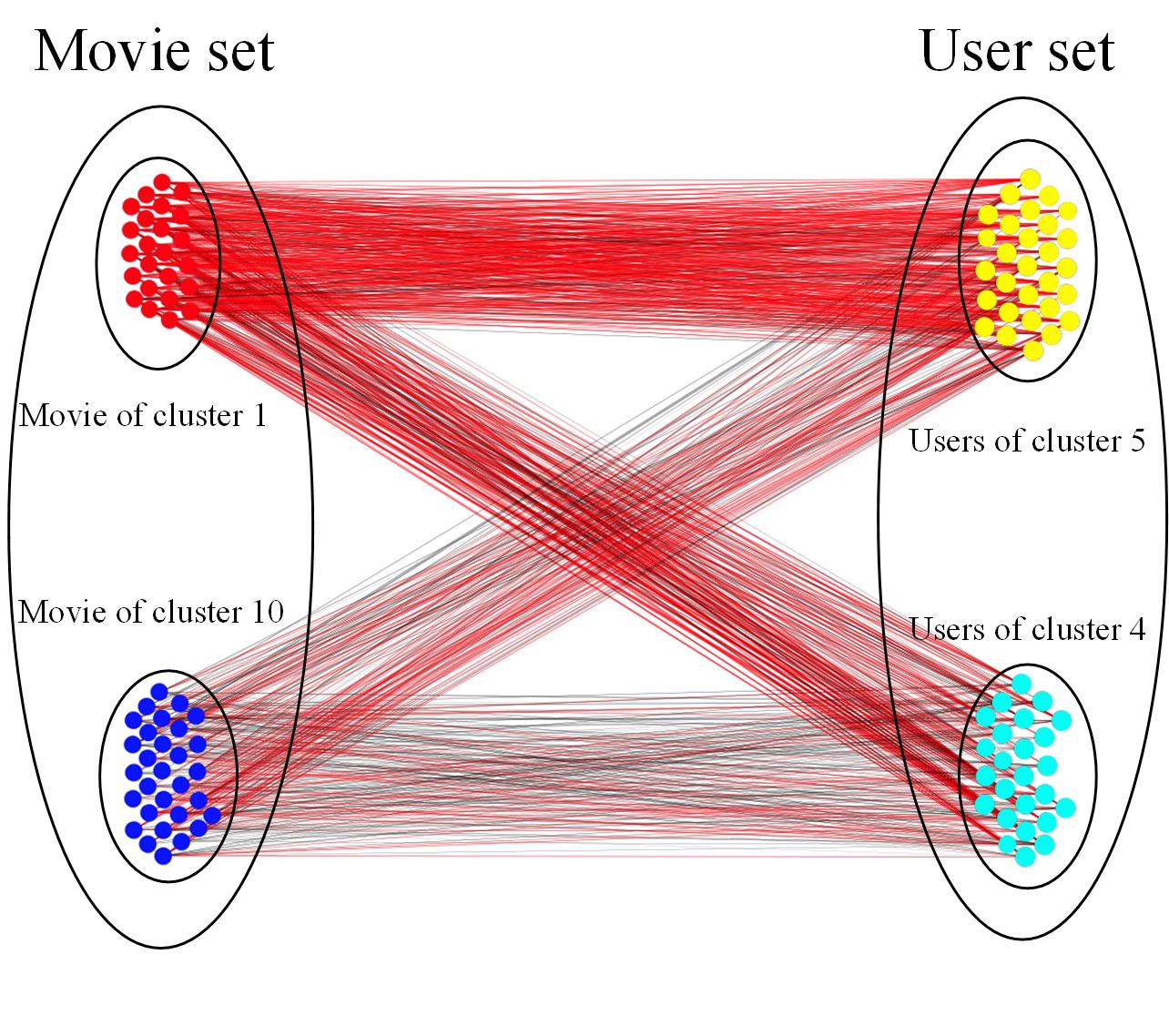}
	\caption{The bipartite network based on the users in 
		cluster $U_4$ and $U_5$ and the movies in cluster $I_1$ and 
		$I_{10}$. Red edges represent high rates, while dark edges 
		represent low rates.}
	\label{fig:movielens_bipartite}
\end{figure}

Figure~\ref{fig:movielens_bipartite} gives the bipartite network 
corresponding to that given in Figure~\ref{fig:movielens_net}. 
For better visualization, we color the rates of $4$ and $5$ with 
red, but those lower rates ($1$, $2$ or $3$) with black. From 
Figure~\ref{fig:movielens_bipartite}, we see most black edges 
emerging between $U_{10}$ and $I_4$, but the majority of the 
edges between $U_{5}$ and $I_{1}$ is red. Thus, the users from 
$U_4$ and $U_5$ show very opposite rating preferences. 
Figure~\ref{fig:movielens_bipartite} also suggests that both 
strict and generous raters tend to rate high-quality movies with 
high scores, but low-quality movies with low scores. Moreover, 
based on edge density, the users from $I_5$ tend to rate more 
movies from $I_1$ than those from $I_{10}$. Even for the users 
from $I_4$, they give more ratings to the movies from $I_1$, 
too. 

\section{Discussions}
\label{sec:dis}

In this paper, we propose an effective model, namely 
$\mathrm{BM}^2$, for 
predicting users' preference in a recommender system. The 
proposed 
model is based on a solid theoretical foundation, where the 
estimation is primarily done in a Bayesian framework. The 
variational 
inference of the model is explicitly discussed, and a 
variational EM 
algorithm is introduced to overcome the potential challenge of 
computational feasibility for massive networks. The proposed 
method remedies the over-fitting problem usually emerging in 
classical collaborative filtering 
approaches~\citep{Salakhutdinov2008bayesian}. 
Additionally, we show by simulations that $\mathrm{BM}^2$ is 
capable of 
handling outliers in contrast to classical SBM or MMSBM. We also 
apply the proposed method to the MovieLens dataset, and find 
that it 
outperforms the competing methods.

The selection of user and item cluster numbers is based on a 
cross-validation method in the present study. One of our future 
study directions is to develop a more rigorous method for 
cluster 
number selection. More generally, we would like to look into 
appropriate model selection criteria like Integrated Complete 
Likelihood~\citep[ICL,][]{Latouche2012variational}. Additionally, 
the present research does not account for missing data, as we assume 
those unobserved data are simply due to the lack interaction between 
the users and items. This assumption may be questionable as there is 
a possibility that the users may refuse to provide the ratings to 
some items for various reasons. Consequently, ignoring these missing 
data potentially causes bias. We will also conduct further 
investigations in this direction in our 
future work.

\section*{Acknowledgments}

We would like to thank two anonymous reviewers and the handling AE 
for insightful comments and suggestions that significantly help 
improve the quality of the paper.

\appendix

\section{Proof of Propositions}
\label{app:propproof}

\subsection{Proof of Proposition~\ref{prop:1}}
\label{app:prop1} 
According to the mean-field variational theory~\citep[Chapter 
10]{Bishop2006pattern} and the variational inference 
algorithm~\citep[Section 2.4]{Blei2017variational}, the 
variational distributions 
$q(\bmpi_{i}^{U} \given \bmgamma_{i}^{U})$ and
$q(\bmpi_{j}^{I} \given \bmgamma_{i}^{I})$ are respectively 
given by
\begin{align*}
	q(\bmpi_{i}^{U} \given \bmgamma_{i}^{U}) & \propto \exp 
	\left\{ \mathbb{E}_{-\bmpi_{i}^{U}} \left[ \log
	p(\bmR,\bmpi,\bmZ \given
	\bmTheta, \bmmu )\right]  \right\} \\
	&\propto \exp \left\{ \mathbb{E}_{- \bmpi_{i}^{U}} \left[
	\log p(\bmpi_{i}^{U} \given \bmalpha ) + \sum_{j \in 
		U_{i}}\log	p(\bmZ_{\itoj}^{U} \given \bmpi_{i}^{U})\right] 
	\right\} \\
	& \propto \exp \left\{ \sum_{k = 1}^{K}\left(\alpha_{k}+ 
	\sum_{j	\in U_{i}}\mathbb{E}\left[\bmZ_{\itoj}^{U}\right] 
	- 1  \right)\log \pi_{ik}^{U}  \right\} \\
	& \sim {\rm Dirichlet}(\bmpi_{i}^{U}; \bmgamma_{i}^{U}),
\end{align*}
and
\begin{align*}
	q(\bmpi_{j}^{I} \given \bmgamma_{j}^{I}) & \propto \exp 
	\left\{	\mathbb{E}_{-\bmpi_{j}^{I}} \left[ \log
	p(\bmR,\bmpi,\bmZ \given \bmTheta, \bmmu )\right]  \right\}\\
	&\propto \exp \left\{ \mathbb{E}_{- \bmpi_{j}^{I}} \left[
	\log p(\bmpi_{j}^{I} \given \beta ) + \sum_{i \in I_{j}}\log
	p(\bmZ_{\ifromj}^{I} \given \bmpi_{j}^{I})\right] \right\}\ 
	\\
	& \propto \exp \left\{ \sum_{l = 1}^{L}(\beta_{l}+ \sum_{i 
		\in	I_{j}}\mathbb{E}\left[\bmZ_{\ifromj}^{I}\right] -1
	)\log \pi_{jl}^{I}  \right\}\\
	& \sim {\rm Dirichlet}(\bmpi_{j}^{I}; \bmgamma_{j}^{I}),
\end{align*}
where $ \gamma_{ik}^{U} = \alpha_{k} + \sum_{j \in U_{i}} 
\phi_{\itoj, k}^{U}$ and $ \gamma_{jl}^{I} = \beta_{l} + \sum_{i 
	\in I_{j}} \phi_{\ifromj, l} $.

\subsection{Proof of Proposition~\ref{prop:2}}
\label{app:prop2} 
The proof is similar to that of Proposition~\ref{app:prop1}. We 
reapply the mean-field variational 
theory~\citep{Bishop2006pattern, Blei2017variational}, and get 
the variational distributions $q(\bmZ_{\itoj}^{U} \given 
\bmphi_{\itoj}^{U})$ and $q(\bmZ_{\ifromj}^{I} \given 
\bmphi_{\ifromj}^{I})$, which are respectively given by

\begin{align*}
	q(\bmZ_{\itoj}^{U} \given \bmphi_{\itoj}^{U}) &\propto 
	\exp\left\{ \mathbb{E}_{-\bmZ_{\itoj}^{U}}\left[ \log
	p(\bmR,\bmpi,\bmZ \given \bmTheta, \boldsymbol{\mu}) \right] 
	\right\} \\		
	&\propto \exp \left\{ \mathbb{E}_{-\bmZ_{\itoj}^{U}}
	\left[  \log p(\bmZ_{\itoj}^{U} \given \bmpi_{i}^{U}) +
	\log p(R_{ij} \given \bmmu, 
	\bmZ_{\itoj}^{U},\bmZ_{\itoj}^{I})  \right] \right\}\\
	&\propto \exp \left\{ \sum_{k = 1}^{K} Z_{\itoj,k}^{U}\left( 
	\mathbb{E}\left[\log \pi_{ik}^{U} \right]
	+ \sum_{l = 1}^{L}\sum_{s = 1}^{S} c_{ij,s} \mathbb{E}\left[
	\bmZ_{\itoj,l}^{I} \right]  \log \mu_{kl,s} \right)
	\right\}\\
	& \sim {\rm 
		Multinomial}(\bmZ_{\itoj}^{U};\bmphi_{\itoj}^{U}),
\end{align*}
and
\begin{align*}
	q(\bmZ_{\ifromj}^{I} \given \bmphi_{\ifromj}^{I}) &
	\propto \exp\left\{ \mathbb{E}_{-\bmZ_{\ifromj}^{I}}\left[ 
	\log p(\bmR,\bmpi,\bmZ \given \bmTheta,\bmmu) \right] 
	\right\}\\
	& \propto \exp \left\{ \mathbb{E}_{-\bmZ_{\ifromj}^{I}}
	\left[ \log p(\bmZ_{\ifromj}^{I} \given 
	\bmphi_{\ifromj}^{I}) + \log p(R_{ij} \given \bmmu, 
	\bmZ_{\itoj}^{U},\bmZ_{\ifromj}^{I})  \right]
	\right\}\\
	& \propto \exp \left\{ \sum_{l = 1}^{L} 
	\bmZ_{\ifromj,l}^{I}\left(\mathbb{E}\left[\log \pi_{jl}^{I} 
	\right] + \sum_{k = 1}^{K}\sum_{s = 1}^{S} 
	c_{ij,s}\mathbb{E}\left[ 
	\bmZ_{\ifromj,l}^{I} \right]  \log \mu_{kl,s}  \right)	
	\right\}\\
	& \sim {\rm Multinomial} 
	(\bmZ_{\ifromj}^{I};\bmphi_{\ifromj}^{I}).
\end{align*}
In addition, by~\citet{Bishop2006pattern}, we have 
$\bmphi^U_{\itoj} := (\phi^{U}_{\itoj, 
	k})_{k = 1}^{K}$ and $\bmphi^I_{\ifromj} := 
(\phi^{I}_{\ifromj,l})_{l = 1}^{L}$	with
\begin{align*}
	\phi^{U}_{\itoj, k} &\propto
	\exp\left\{\psi(\gamma_{ik}^{U}) - \psi\left(\sum_{k =
		1}^{K} \gamma_{ik}^{U}\right) + \sum_{l = 1}^{L}
	\sum_{s = 1}^{S} \bm{1}(R_{ij} = s) \phi_{\ifromj, l}^{I} 
	\log{\mu_{kl, s}}\right\}, \\
	\phi^{I}_{\ifromj, k} &\propto
	\exp\left\{\psi(\gamma_{jl}^{I}) - \psi\left(\sum_{l =
		1}^{L} \gamma_{jl}^{I}\right) + \sum_{k = 1}^{K}
	\sum_{s = 1}^{S} \bm{1}(R_{ij} = s) \phi_{\itoj, k}^{U} 
	\log{\mu_{kl, s}}\right\}.
\end{align*}

\subsection{Proof of Proposition~\ref{prop:3}}
\label{app:prop3} 
According to Equation~\eqref{eq:43}, we can split the variational 
lower bound into the sum of two expectations as follows:
\begin{align*}
	\mathcal{L}(q(\bmpi, \bmZ); \bmmu, \bmTheta) & = \int_{\bmpi}
	\sum_{\bmZ} q(\bmpi, \bmZ) \log \frac{p(\bmpi, \bmZ, \bmR 
		\given
		\bmmu, \bmTheta)}{q(\bmpi, \bmZ)} \, {\rm d}\bmpi\\
	& = \int_{\bmpi} \sum_{\bmZ} q(\bmpi, \bmZ) \log p(\bmpi, \bmZ, 
	\bmR \given \bmmu, \bmTheta)\, {\rm d}\bmpi - \int_{\bmpi} 
	\sum_{\bmZ} q(\bmpi, \bmZ) \log q(\bmpi, \bmZ) \, {\rm d}\bmpi\\
	& =  \mathbb{E}_{q}\left[ \log p(\bmR,\bmpi,\bmZ \given \bmmu, 
	\bmTheta) \right] - \mathbb{E}_{q}\left[  \log	q(\bmpi,\bmZ) 
	\right]\\
	& = I_{1}  - I_{2}.
\end{align*}
We look into $I_1$ and $I_2$ one after another. According to 
Equation~\eqref{eq:model}, $I_{1}$ can be rewritten as:
\begin{align*}
	I_{1} & =  \sum_{i = 1}^{N} \mathbb{E}_{q}\left[\log 
	p(\bmpi_{i}^{U} \given \bmalpha) \right] + \sum_{j = 
		1}^{M}\mathbb{E}_{q}\left[\log p(\bmpi_{j}^{I}\given 
		\bmbeta) 
	\right] + \sum_{R_{ij} \in \bmR} \left(\mathbb{E}_{q}\left[ \log 
	p(\bmZ_{\itoj}^{U} \given \bmpi_{i}^{U})\right]\right. 
	\\ &\qquad\qquad{} + \left. 
	\mathbb{E}_{q} \left[ \log p(\bmZ_{\ifromj}^{I} \given 
	\bmpi_{j}^{I}) \right] \right) + \sum_{R_{ij} \in \bmR} \left( 
	\mathbb{E}_{q} \left[ 
	\log 
	p(\bmc_{ij} \given 
	\bmZ_{\itoj}^{U},\bmZ_{\ifromj}^{I},\bmmu)\right] \right).
\end{align*}
On the other hand, by Equation~\eqref{eq:42}, we divide $I_{2}$ into 
the following parts:
\begin{align*}
	I_{2}  &=  \sum_{i = 1}^{N} \mathbb{E}_{q} \left[\log 
	q(\bmpi_{i}^{U}	\given \bmgamma_{i}^{U}) \right] + \sum_{j 
		= 1}^{M}\mathbb{E}_{q} \left[\log q(\bmpi_{j}^{I} \given 
	\bmgamma_{j}^{I}) \right] +\sum_{R_{ij}\in \bmR} \left(
	\mathbb{E}_{q}\left[ \log q(\bmZ_{\itoj}^{U} \given
	\bmphi_{\itoj}^{U}) \right]\right. 
	\\&\qquad\qquad{}+ \left. \mathbb{E}_{q}
	\left[\log q(\bmZ_{\ifromj}^{I} \given \bmphi_{\ifromj}^{I}) 
	\right] \right).
\end{align*}

By Propositions~\ref{prop:1} and ~\ref{prop:2}, the expectations 
in $I_1$ and $I_2$ are respectively given by
\begin{align*}
	&\mathbb{E}_{q} \left[ \log \pi^{U}_{ik} \right] =
	\psi(\gamma_{ik}^{U}) - \psi(\sum_{k=1}^{K}
	\gamma_{ik}^{U}) \qquad {\rm and} \qquad
	\mathbb{E}_{q} \left[ \log \pi^{I}_{jl} \right] =
	\psi(\gamma_{jl}^{I}) - \psi(\sum_{l=1}^{L}
	\gamma_{jl}^{I}),\\
	&\mathbb{E}_{q} \left[ Z_{\itoj,k}^{U}\right] =
	\phi_{\itoj,k}^{U} \qquad {\rm and} \qquad
	\mathbb{E}_{q} \left[ Z_{\itoj,l}^{I}\right] =
	\phi_{\itoj,l}^{I}.\\
\end{align*}

For the first two terms in $I_{1}$, we calculate the expectation 
of $\bmpi_{i}^{U}$ 
with respect to variational distribution
$q(\bmpi,\bmZ)$ as follows

\begin{align*}
	\mathbb{E}_{q}\left[ \log p(\bmpi_{i}^{U} \given \bmalpha) 
	\right]
	& = \sum_{k=1}^{K} (\alpha_{k} - 1)\mathbb{E}_{q} \left[\log
	\pi_{ik}^{U} \right] + f_{1}(\bmalpha)\\
	& = \sum_{k=1}^{K} (\alpha_{k} - 
	1)\left(\psi(\gamma_{ik}^{U}) -
	\psi\left(\sum_{k=1}^{K}\gamma_{ik}^{U}\right)\right) +  
	f_{1}(\bmalpha )\\
	& = \sum_{k=1}^{K} (\alpha_{k} - 
	1)f_{2}\left(\gamma_{ik}^{U},\bmgamma_{i}^{U}\right) +  
	f_{1}(\bmalpha ),
\end{align*}
where $f_1(\bmx) = \log \Gamma\left(\sum_{d = 1}^{D} x_d\right) -
\sum_{d = 1}^{D} \log \Gamma(x_d)$ and $ f_2(x_d, \bmx) = 
\psi(x_d) - \psi\left(\sum_{d = 1}^{D}
x_d\right)$. The interpretation of $f_1(\bmx)$ is the logarithm 
of the constant for Dirichlet distribution, and $f_2(x_d, \bmx)$ 
is the expectation of the logarithm of the dimension of $\bmx$. 
We exploit an analogous argument to compute 
$\mathbb{E}_{q} \left[\log q(\bmpi_{j}^{I} \given 
\bmgamma_{j}^{I}) \right]$, and omit the details.

For the next two terms in $I_{1}$, we only present the 
conditional expectation of $\bmZ_{\itoj}^{U}$, as the 
conditional expectation of $\bmZ_{\ifromj}^{I}$ can be obtained 
in a similar manner.	
\begin{align*}
	\mathbb{E}_{q} \left[ \log p(\bmZ_{\itoj}^{U}\mid
	\bmpi_{i}^{U}) \right] & = \sum_{k=1}^{K} \mathbb{E}_{q}
	\left[ Z_{\itoj,k}^{U} \right] \mathbb{E}_{q} \left[
	\log \pi_{ik}^{U} \right]\\
	& = \sum_{k=1}^{K} \phi_{\itoj,k}
	(\psi(\gamma_{ik}^{U}) - \psi(\sum_{k=1}^{K}\gamma_{ik}^{U}))\\
	& = \sum_{k=1}^{K} \phi_{\itoj,k} 
	f_{2}(\gamma_{ik}^{U},\bmgamma_{i}^{U})
\end{align*}

For the last term in $I_{1}$, we have
\begin{align*}
	\mathbb{E}_{q} \left[ \log p(\bmc_{ij} \given 
	\bmZ_{\itoj}^{U},\bmZ_{\ifromj}^{I},\bmmu)\right] & =
	\sum_{k=1}^{K} \sum_{l=1}^{L} \mathbb{E}_{q} \left[ 
	Z_{\itoj,k}^{U} \right] \mathbb{E}_{q} 
	\left[Z_{\ifromj,l}^{I} \right] \sum_{s=1}^{S} c_{ij,s} \log
	\mu_{kl,s}\\
	& = \sum_{k=1}^{K} \sum_{l=1}^{L} \sum_{s=1}^{S} 
	\phi_{\itoj,k} \phi_{\ifromj,l}c_{ij,s}\log
	\mu_{kl,s}.
\end{align*}

For the first two terms in $I_{2}$, we present the expectation 
of $\log q(\bmpi_{i}^{U} \given	\bmgamma_{i}^{U})$ only, since 
the other one can be done similarly.
\begin{align*}
	\mathbb{E}_{q} \left[ \log q(\bmpi_{i}^{U} \mid
	\bmgamma_{i}^{U}) \right] &= \sum_{k=1}^{K}
	(\gamma_{ik}^{U}-1)\mathbb{E}_{q} \left[ \log
	\pi_{ik}^{U}\right] +  f_{1}(\bmgamma_{i}^{U})\\
	& = \sum_{k = 1}^{K} (\gamma_{ik}^{U}-1)\left(
	\psi(\gamma_{ik}^{U}) - \psi\left(\sum_{k=1}^{K}
	\gamma_{ik}^{U}\right)\right) + f_{1}(\bmgamma_{i}^{U}) \\ 
	& = \sum_{k = 1}^{K} 
	(\gamma_{ik}^{U}-1)f_{2}\left(\gamma_{ik}^{U},
	\bmgamma_{i}^{U}\right) + f_{1}(\bmgamma_{i}^{U}).
\end{align*}

For the third and fourth terms in $I_{2}$, we only need to show 
the conditional expectation of $\log q(\bmZ_{\itoj}^{U} \given	
\bmphi_{\itoj}^{U})$.
\begin{align*}
	\mathbb{E}_{q}\left[ \log q(\bmZ_{\itoj}^{U} \mid
	\bmphi_{\itoj}^{U})\right] & = \sum_{k=1}^{K} 
	\mathbb{E}_{q}\left[
	Z_{\itoj,k}^{U} \right] \log \phi_{\itoj,k}^{U} \\
	& = \sum_{k=1}^{K} \phi_{\itoj,k}^{U} \log \phi_{\itoj}^{U}.
\end{align*}
The proof is completed by putting all the expectations together.

\subsection{Proof of Proposition~\ref{prop:4}}
\label{app:prop4} 

We maximize the ELBO, i.e., 
$\mathcal{L}(q(\bm{\pi},\bm{Z}); \bmmu,\bmTheta)$
subject to the constraints $\sum_{q=1}^{Q} \mu_{kl,q} = 1,
k=1,\dots,K; l=1,\ldots,L$ by applying the Lagrange multiplier 
method as follows:
\begin{align*}
	L(\bm{\mu},\lambda) = \sum_{R_{ij} \in \bmR} \sum_{k=1}^{K} 
	\sum_{l=1}^{L}\sum_{s=1}^{S}
	\phi_{\itoj,k}^{U} \phi_{\ifromj,l}^{I}
	c_{ij,s} \log{\mu_{kl,s}} + \sum_{k=1}^{K} \sum_{l=1}^{L}
	\lambda_{kl}(\sum_{s=1}^{S} \mu_{kl,s} - 1).
\end{align*}

Take the partial derivatives with respect to $\mu_{kl,q}$ and
$\lambda_{kl}$:
\begin{align*}
	\frac{\partial L}{ \partial \mu_{kl,s}} &= \sum_{R_{ij} \in
		\bmR} \phi_{\itoj,k}^{U} \phi_{\ifromj,l}^{I} 
	c_{ij,s}\frac{1}{\mu_{kl,q}} +
	\lambda_{kl}, \\
	\frac{\partial L}{\lambda_{kl}} &= \sum_{s=1}^{S} \mu_{kl,s}
	-1, \\
	\frac{\partial^{2} L}{ \partial \mu^{2}_{kl,s}} &=
	-\frac{1}{\mu^{2}_{kl,s}} \sum_{R_{ij} \in \bmR}
	\phi_{\itoj,k}^{U} \phi_{\ifromj,l}^{I}
	c_{ij,s} < 0.  \\
\end{align*}
Set the first order partial	derivatives to equal zero, and we
can get the closed form for the solution of $\bmmu$.
\begin{align*}
	\mu_{kl,s} = \frac{\sum_{R_{ij} \in 
			\bmR}\phi_{\itoj,k}^{U}\phi_{\ifromj,l}^{I}
		R_{ij}\bm{1}(R_{ij} = s)}{\sum_{R_{ij} \in 
			\bmR}\sum_{s=1}^{S}
		\phi_{\itoj,k}^{U} \phi_{\ifromj,l}^{I}}
\end{align*}

\section{Parameters in simulation}
\label{app:simmu}

We give the block matrices $\bmmu_{1}$, $\bmmu_{2}$, 
$\bmmu_{3}$, $\bmmu_{4}$ and $\bmmu_{5}$ that are used for 
the simulations in Section~\ref{sec:sim} as follows.
For $K = L = 5$,	
\begin{align*}
	\bmmu_1 &:= \begin{pmatrix}
		0.65 & 0.45 & 0.25 & 0.15 & 0.10\\
		0.45 & 0.25 & 0.05 & 0.05 & 0.05\\
		0.10 & 0.10 & 0.05 & 0.05 & 0.05\\
		0.10 & 0.10 & 0.02 & 0.02 & 0.02\\
		0.10 & 0.12 & 0.02 & 0.02 & 0.02
	\end{pmatrix},\\
	\\ \bmmu_2 &:= \begin{pmatrix}
		0.18 & 0.28 & 0.38 & 0.28 & 0.25 \\
		0.28 & 0.38 & 0.48 & 0.30 & 0.20 \\
		0.40 & 0.20 & 0.10 & 0.10 & 0.10 \\
		0.35 & 0.20 & 0.05 & 0.05 & 0.05 \\
		0.25 & 0.15 & 0.05 & 0.05 & 0.05
	\end{pmatrix},\\
	\\ \bmmu_3 &:= \begin{pmatrix}
		0.10 & 0.28 & 0.30 & 0.30 & 0.30 \\
		0.20 & 0.30 & 0.40 & 0.30 & 0.30 \\
		0.35 & 0.45 & 0.50 & 0.30 & 0.20 \\
		0.30 & 0.35 & 0.40 & 0.30 & 0.20 \\
		0.30 & 0.30 & 0.30 & 0.20 & 0.10
	\end{pmatrix},\\
	\\ \bmmu_4 &:= \begin{pmatrix}
		0.05 & 0.05 & 0.05 & 0.25 & 0.25 \\
		0.05 & 0.05 & 0.05 & 0.25 & 0.35 \\
		0.10 & 0.20 & 0.30 & 0.40 & 0.40 \\
		0.20 & 0.30 & 0.48 & 0.38 & 0.28 \\
		0.25 & 0.28 & 0.38 & 0.28 & 0.18
	\end{pmatrix},\\
	\\ \bmmu_5 &:= \begin{pmatrix}
		0.02 & 0.02 & 0.02 & 0.10 & 0.10 \\
		0.02 & 0.02 & 0.02 & 0.10 & 0.10 \\
		0.05 & 0.05 & 0.05 & 0.15 & 0.25 \\
		0.05 & 0.05 & 0.05 & 0.25 & 0.45 \\
		0.10 & 0.15 & 0.25 & 0.45 & 0.65
	\end{pmatrix}.
\end{align*}

For $K = L = 7$,

\begin{align*}
	\bmmu_1 &:= \begin{pmatrix}
		0.65  &  0.55  &  0.45  &  0.35  &  0.25  &  0.15  &  0.10\\
		0.45  &  0.35  &  0.25  &  0.15  &  0.05  &  0.05  &  0.05\\
		0.25  &  0.20  &  0.15  &  0.10  &  0.05  &  0.05  &  0.05\\
		0.10  &  0.10  &  0.10  &  0.05  &  0.05  &  0.05  &  0.05\\
		0.10  &  0.10  &  0.10  &  0.05  &  0.05  &  0.05  &  0.05\\
		0.10  &  0.10  &  0.02  &  0.02  &  0.02  &  0.02  &  0.02\\
		0.10  &  0.10  &  0.06  &  0.06  &  0.02  &  0.02  &  0.02
	\end{pmatrix},\\
	\\ \bmmu_2 &:= \begin{pmatrix}
		0.18  &  0.23  &  0.28  &  0.33  &  0.38  &  0.28  &  0.25\\
		0.28  &  0.33  &  0.38  &  0.43  &  0.48  &  0.30  &  0.25\\
		0.45  &  0.45  &  0.40  &  0.35  &  0.30  &  0.25  &  0.20\\
		0.40  &  0.30  &  0.20  &  0.10  &  0.10  &  0.10  &  0.10\\
		0.35  &  0.25  &  0.15  &  0.05  &  0.05  &  0.05  &  0.10\\
		0.35  &  0.20  &  0.05  &  0.05  &  0.05  &  0.05  &  0.05\\
		0.25  &  0.20  &  0.15  &  0.10  &  0.05  &  0.05  &  0.05
	\end{pmatrix},\\
	\\ \bmmu_3 &:= \begin{pmatrix}
		0.10  &  0.15  &  0.20  &  0.25  &  0.30  &  0.30  &  0.30\\
		0.20  &  0.25  &  0.30  &  0.35  &  0.40  &  0.30  &  0.35\\
		0.15  &  0.20  &  0.30  &  0.40  &  0.45  &  0.40  &  0.35\\
		0.35  &  0.40  &  0.45  &  0.50  &  0.40  &  0.30  &  0.20\\
		0.40  &  0.45  &  0.40  &  0.40  &  0.35  &  0.30  &  0.25\\
		0.30  &  0.35  &  0.40  &  0.35  &  0.30  &  0.25  &  0.20\\
		0.30  &  0.35  &  0.35  &  0.30  &  0.30  &  0.20  &  0.10
	\end{pmatrix},
\end{align*}
\begin{align*}
	\bmmu_4 &:= \begin{pmatrix}
		0.05  &  0.05  &  0.05  &  0.05  &  0.05  &  0.20  &  0.25\\
		0.05  &  0.05  &  0.05  &  0.05  &  0.05  &  0.25  &  0.25\\
		0.10  &  0.10  &  0.10  &  0.10  &  0.15  &  0.20  &  0.30\\
		0.10  &  0.15  &  0.20  &  0.30  &  0.35  &  0.40  &  0.40\\
		0.10  &  0.15  &  0.30  &  0.40  &  0.45  &  0.40  &  0.40\\
		0.20  &  0.30  &  0.48  &  0.43  &  0.38  &  0.33  &  0.28\\
		0.25  &  0.25  &  0.29  &  0.34  &  0.38  &  0.28  &  0.18
	\end{pmatrix},\\
	\\ \bmmu_5 &:= \begin{pmatrix}
		0.02  &  0.02  &  0.02  &  0.02  &  0.02  &  0.07  &  0.10\\
		0.02  &  0.02  &  0.02  &  0.02  &  0.02  &  0.10  &  0.10\\
		0.05  &  0.05  &  0.05  &  0.05  &  0.05  &  0.10  &  0.10\\
		0.05  &  0.05  &  0.05  &  0.05  &  0.10  &  0.15  &  0.25\\
		0.05  &  0.05  &  0.05  &  0.10  &  0.10  &  0.20  &  0.20\\
		0.05  &  0.05  &  0.05  &  0.15  &  0.25  &  0.35  &  0.45\\
		0.10  &  0.10  &  0.15  &  0.20  &  0.25  &  0.45  &  0.65
	\end{pmatrix}.
\end{align*}

For $K = L = 9$,
\[	\bmmu_1 := \begin{pmatrix}
	0.70  &  0.65  &  0.55  &  0.45  &  0.35  &  0.25  &  0.15  &  
	0.10  &  0.10\\
	0.55  &  0.45  &  0.35  &  0.25  &  0.15  &  0.05  &  0.05  &  
	0.05  &  0.05\\
	0.40  &  0.30  &  0.20  &  0.10  &  0.10  &  0.05  &  0.05  &  
	0.05  &  0.05\\
	0.25  &  0.15  &  0.05  &  0.05  &  0.05  &  0.05  &  0.05  &  
	0.05  &  0.05\\
	0.10  &  0.10  &  0.10  &  0.10  &  0.05  &  0.05  &  0.05  &  
	0.05  &  0.05\\
	0.10  &  0.10  &  0.10  &  0.02  & 0.02   & 0.02   & 0.02   & 
	0.02   & 0.02\\
	0.10  &  0.10  &  0.02  &  0.02  &  0.02  &  0.02  &  0.02  &  
	0.02  &  0.02\\
	0.10  &  0.10  &  0.10  &  0.06  &  0.06  &  0.02  &  0.02  &  
	0.02  &  0.02\\
	0.05  &  0.05  &  0.05  &  0.02  &  0.02  &  0.02  &  0.02  &  
	0.02  &  0.02
\end{pmatrix},\]
\begin{align*}
	\bmmu_2 &:= \begin{pmatrix}
		0.15  &  0.18  &  0.23  &  0.28  &  0.33  &  0.38  &  0.28  
		&  0.25  &  0.20\\
		0.23  &  0.28  &  0.33  &  0.38  &  0.43  &  0.48  &  0.30  
		&  0.20  &  0.15\\
		0.31  &  0.36  &  0.41  &  0.46  &  0.36  &  0.31  &  0.21  
		&  0.21  &  0.21\\
		0.41  &  0.41  &  0.41  &  0.36  &  0.31  &  0.26  &  0.16  
		&  0.11  &  0.11\\
		0.50  &  0.40  &  0.30  &  0.20  &  0.10  &  0.10  &  0.10  
		&  0.10  &  0.10\\
		0.40  &  0.35  &  0.20  &  0.05  &  0.05  &  0.05  &  0.05  
		&  0.05  &  0.05\\
		0.35  &  0.20  &  0.05  &  0.05  &  0.05  &  0.05  &  0.05  
		& 0.05   & 0.05\\
		0.30  & 0.25   & 0.20   & 0.15   & 0.10   & 0.05   & 0.05   
		& 0.05   & 0.05\\
		0.25  &  0.15  &  0.15  &  0.08  &  0.08  &  0.08  &  0.03  
		&  0.03  &  0.03\\
	\end{pmatrix},\\
	\\ \bmmu_3 &:= \begin{pmatrix}
		0.08  &  0.10  &  0.15  &  0.20  &  0.25  &  0.30  &  0.30  
		&  0.30  &  0.30\\
		0.15  &  0.20  &  0.25  &  0.30  &  0.35  &  0.40  &  0.30  
		&  0.30  &  0.35\\
		0.22  &  0.27  &  0.32  &  0.37  &  0.42  &  0.47  &  0.42  
		&  0.32  &  0.27\\
		0.27  &  0.32  &  0.37  &  0.42  &  0.47  &  0.52  &  0.42  
		&  0.32  &  0.27\\
		0.30  &  0.35  &  0.40  &  0.45  &  0.50  &  0.40  &  0.30  
		&  0.20  &  0.15\\
		0.35  &  0.30  &  0.35  &  0.40  &  0.35  &  0.30  &  0.25  
		&  0.20  &  0.15\\
		0.30  &  0.35  &  0.40  &  0.35  &  0.30  &  0.25  &  0.20  
		&  0.15  &  0.10\\
		0.30  &  0.30  &  0.35  &  0.35  &  0.30  &  0.30  &  0.20  
		&  0.10  &  0.10\\
		0.40  &  0.50  &  0.40  &  0.30  &  0.30  &  0.20  &  0.10  
		&  0.05  &  0.05		
	\end{pmatrix},\\
	\\ \bmmu_4 &:= \begin{pmatrix}
		0.05  &  0.05  &  0.05 &   0.05  &  0.05 &   0.05  &  0.20  
		&  0.25  &  0.30\\
		0.05  &  0.05  &  0.05 &   0.05  &  0.05 &   0.05  &  0.25  
		&  0.35  &  0.35\\
		0.05  &  0.05  &  0.15 &   0.15  &  0.15 &   0.15  &  0.30  
		&  0.35  &  0.40\\
		0.05  &  0.10  &  0.15 &   0.20  &  0.30 &   0.35  &  0.40  
		&  0.40  &  0.40\\
		0.10  &  0.20  &  0.30 &   0.48  &  0.43 &   0.38  &  0.33  
		&  0.28  &  0.28\\
		0.20  &  0.30  &  0.48 &   0.43  &  0.38 &   0.33  &  0.28  
		&  0.28  &  0.28\\
		0.20  &  0.25  &  0.25 &   0.29  &  0.34 &   0.38  &  0.28  
		&  0.18  &  0.13\\
		0.20  &  0.20  &  0.30 &   0.40  &  0.30 &  0.40   & 0.40   
		&  0.25  &  0.15
	\end{pmatrix},
\end{align*}
\[ \bmmu_5 := \begin{pmatrix}
	0.02  &  0.02  &  0.02  &  0.02  &  0.02  &  0.02   & 0.07  &  
	0.10  &  0.10\\
	0.02  &  0.02  &  0.02  &  0.02  &  0.02  &  0.02   & 0.10  &  
	0.10  &  0.10\\
	0.02  &  0.02  &  0.02  &  0.02  & 0.02   & 0.02   & 0.07   & 
	0.07   & 0.07\\
	0.02  &  0.02  &  0.02  &  0.02  &  0.02  &  0.02   & 0.07  &  
	0.17  &  0.17\\
	0.05  &  0.05  &  0.05  &  0.05  &  0.05   & 0.10   & 0.15  &  
	0.25  &  0.30\\
	0.05  &  0.05  &  0.05  &  0.05  &  0.15   & 0.25   & 0.35  &  
	0.45  &  0.50\\
	0.05  &  0.05  &  0.05  & 0.15   & 0.25    &0.35    & 0.45  &  
	0.50  &  0.55\\
	0.10  &  0.10  &  0.10  &  0.15  &  0.20   & 0.25   & 0.45  &  
	0.65  &  0.70\\
	0.10  &  0.10  &  0.10  &  0.20  &  0.30   & 0.30   & 0.45  &  
	0.65  &  0.75
\end{pmatrix}.
\]
Additionally, we give the true values for $\bmalpha$ and 
$\bmbeta$ in the simulation. For $K=L=5$,
\begin{align*}
	\bmalpha &= (0.10;0.20;0.40;0.20;0.10),\\
	\bmbeta &= (0.10;0.15;0.45;0.25;0.05).
\end{align*}
For $K=L=7$,
\begin{align*}
	\bmalpha &= (0.07,0.11,0.17,0.30,0.17,0.11,0.07),\\
	\bmbeta &= (0.07,0.10,0.15,0.41,0.14,0.08,0.05).
\end{align*}
For $K = L = 9$,
\begin{align*}
	\bmalpha &= (0.02,0.05,0.12,0.16,0.30,0.12,0.15,0.05,0.03)\\
	\bmbeta &= (0.03,0.05,0.12,0.17,0.30,0.14,0.12,0.07,0.03).
\end{align*}	



	
	
	
\end{document}